\documentclass[12pt]{article}

\usepackage{amsmath}
\usepackage[english]{babel}
\usepackage[ansinew]{inputenc}
\usepackage{epsfig}
\usepackage{color,graphicx}
\usepackage{wrapfig}
\usepackage{amssymb,amsthm,amsmath}
\usepackage{dsfont}
\usepackage{natbib}
\usepackage{abstract}

\usepackage{enumerate}
\usepackage{multirow}

\newcommand{\blind}{1}

\renewcommand{\baselinestretch}{1}

\newtheorem{theorem}{Theorem}[]
\newtheorem{lemma}{Lemma}[]

\newtheorem{proposition}{Proposition}[]
\newtheorem{corollary}{Corollary}[]

\newtheorem*{thma}{Proof}

\def\ds{\displaystyle}

\begin{document}

\def\spacingset#1{\renewcommand{\baselinestretch}%
{#1}\small\normalsize} \spacingset{1}


\if1\blind
{
  \title{\bf Exact Bayesian inference for diffusion-driven Cox processes}
  \author{Fl\'{a}vio B. Gonçalves\\
    Universidade Federal de Minas Gerais, Brazil\\
    \\
    Krzysztof G. {\L}atuszy\'{nski}\\
    University of Warwick, UK
    and \\
    \\
    Gareth O. Roberts\\
    University of Warwick, UK}
  \maketitle
} \fi

\if0\blind
{
  \bigskip
  \bigskip
  \bigskip
  \begin{center}
    {\LARGE\bf Exact Bayesian inference for diffusion-driven Cox processes}
\end{center}
  \medskip
} \fi

\bigskip
\begin{abstract}
In this paper, we present a novel methodology to perform Bayesian inference for Cox processes in which the intensity function is driven by a diffusion process. The novelty lies in the fact that no discretization error is involved, despite the non-tractability of both the likelihood function and the transition density of the diffusion. The methodology is based on an MCMC algorithm and its exactness is built on retrospective sampling techniques. The efficiency of the methodology is investigated in some simulated examples and its applicability is illustrated in some real data analyzes.
\end{abstract}

\noindent%
\vfill

\newpage
\spacingset{1} 

\section{Introduction}

A Cox process (also sometimes termed doubly stochastic Poisson process) is a Poisson process in which the intensity function (IF) evolves stochastically. Cox processes \citep{Cox2} have been extensively used in a variety of areas to model point process phenomena. Examples can be found in finance - to model credit risk \citep{cps,cariboni}, survival analysis \citep{sangalli}, internet traffic \citep{iversen}, insurance \citep{dassios} and biology \citep{legg}.

We consider unidimensional Cox processes which models the evolution of the IF by means of a diffusion process. We call the resulting process a \emph{diffusion-driven Cox process} (DDCP). A diffusion process is a continuous time (univariate) Markov process which is defined as the solution of a stochastic differential equation (SDE) of the type:
\begin{equation}\label{sde}
  dY_s = a(Y_s;\theta)ds+\sigma(Y_s;\theta)dW_s,\;\;\;Y_0\sim f_{0}^*,
\end{equation}
where $W_s$ is a Brownian motion and $a:\mathds{R}\rightarrow\mathds{R}$ and $\sigma:\mathds{R}\rightarrow\mathds{R}^+$ are assumed to satisfy the regularity conditions (locally Lipschitz, with a linear growth bound) to guarantee a unique weak solution \citep[see][Chapter 4]{kloeden}. Somewhat more general diffusion processes (eg. time inhomogeneous and multivariate) can be dealt with within the framework we provide.
For an accessible introduction to SDEs, see \citet{oksendal}.

Compared to the most popular unidimensional non-parametric Cox processes in which the IF is a function of a Gaussian process, DDCP offers a range of new possibilities to model the stochastic dynamics of the IF. This follows a common general direction taken with other classes of statistical models in which more flexible structures are proposed for model components traditionally assumed to be Gaussian.

Suppose $N:=\{N_s; s\in[0,T]\}$ is an one-dimensional inhomogeneous Poisson process (PP), with intensity function $\lambda:=\{\lambda_s; s\in[0,T]\}$, observed in a time interval $[0,T]$.
We consider DDCPs of the type:
\begin{eqnarray}
\ds  N    &\sim& PP(\lambda_s),\;\;\;s\in[0,T]; \label{DDCPeq1}\\
  \lambda_s  &=& g(X_s;\theta); \label{DDCPeq2}\\
  dX_s &=& \alpha(X_s;\theta)ds+dW_s; \label{DDCPeq3}\\
  X_0 &\sim& f_{0}(\cdot;\theta).\label{DDCPeq4}
\end{eqnarray}
The IF of the Poisson process $N$ is a function $g$ of a diffusion process $X:=\{X_s; s\in[0,T]\}$, where $g(\cdot;\theta):\mathds{R}\rightarrow\mathds{R}^+$ is non-negative and non-explosive, and $f_{0}$ is the Lebesgue density of $X_0$. The diffusion drift $\alpha$ is presumed to satisfy the regularity conditions (locally Lipschitz, with a linear growth bound) that guarantee the existence of a weakly unique, global solution of the SDE. $\theta$ is a vector of unknown parameters. One may choose different parametrizations of the model by manipulating the dependency of $g$ and $X$ on $\theta$. The parametrization should be chosen taking into account the interpretation of the model and its impact on the inference methodology - to be discussed further in Section \ref{ssecparam}. Finally, note that we are not restricted to unit diffusion coefficient diffusions. As long as the coefficient $\sigma$ in (\ref{sde}) is continuously differentiable, we can rewrite a chosen intensity function $h(Y)$ as $g(X)$, where $g=h\circ\eta^{-1}$ and $\eta(y,\theta)=\ds \int_{y^*}^{y}\frac{1}{\sigma(u;\theta)}du$, the Lamperti transform of $Y$, for some $y^*$ in the state space of $Y$.

As is common in computational Bayesian methodologies involving intractable likelihoods, carrying out
inference under the model in (\ref{DDCPeq1})-(\ref{DDCPeq4})
is closely linked to being able to simulate from the model, which is itself a particularly challenging problem. As a result of this,
existing approaches to this problem \citep[see, for example,][]{cps,cariboni,lech} have resorted to
discrete time approximations, often leading to significant (and typically difficult to quantify) bias as well as substantial computational overhead.

The aim of this paper is to propose a methodology that is free of discretization error to perform simulation and inference for DDCPs as in (\ref{DDCPeq1})-(\ref{DDCPeq4}). We term the methodology as exact in the sense of  Monte Carlo error and MCMC convergence are the only sources of approximation. The proposed methodology consists of an MCMC algorithm to sample from the posterior distribution of the unknown components in the model, i.e., parameters and IF. Although the IF is infinite-dimensional, the proposed MCMC is actually based on a finite (albeit varying) dimensional Markov chain. This is due to the retrospective sampling approach adopted, in which the Markov chain contains the unknown parameters of the model and a random finite-dimensional representation of the IF. This representation is such that the algorithm is tractable and the posterior distribution of the remainder of the IF can be easily recovered. Further conditions on functions $\alpha$ and $g$ are required but still consider a wide and flexible range of models. In this context, two particular forms for function $g$ are highlighted, given their good modeling and inference properties. Extensions to consider different data schemes and to more general models are also discussed. The latter is based on recent work on exact inference for jump-diffusions \citep[see][]{GRL} that relies on an infinite-dimensional Barker's MCMC via Bernoulli factories.
Finally, the proposed methodology is investigated in simulated examples and its application is explored with real datasets. In particular, we illustrate the flexibility of the DDCPs when compared to more commonly used Cox process models by considering a diffusion $X$ with a Cauchy invariant distribution. The advantages of the exact approach over discretized ones is explored by comparing the performance of both methodologies in some simulated examples.

The retrospective sampling approach used in this paper is based on previous work on exact inference for discretely observed diffusions \citep[see][]{bpr06a}. Nevertheless, significant differences between the two methodologies, like the fact that for DDCPs the diffusion process is completely latent and plays the role of a non-parametric prior on the IF, require the development of novel and non-trivial simulation techniques and the derivation of novel theoretical results. In particular, the complexity of Poisson process likelihood function makes it considerably harder to: 1. devise a rejection sampling algorithm that samples diffusion bridges from their respective full conditional distributions; 2. obtain the full conditional density of the model parameters $\theta$.

This paper is organized as follows. Section \ref{secinf1} presents the methodology to perform exact Bayesian inference for DDCPs. Simulated examples to investigate the efficiency of the proposed methodology and compare this to dicretization-based approaches are presented in Section \ref{secexamp}. Three real examples, including prediction exercises, are presented in Section \ref{secapp}. Finally, Section \ref{secFT} discusses some further topics including model parametrization, prediction, inference for different data schemes and extensions of the proposed methodology.

\section{Bayesian inference for DDCPs}\label{secinf1}

Consider the DDCP model in (\ref{DDCPeq1})-(\ref{DDCPeq4}) and suppose that $N$ is observed in $[0,T]$. Define $\{t_1,t_2,\ldots,t_n\}$ as the $n$ observed events from $N$ in $[0,T]$, i.e., the dataset based on which inference is to be performed. Our aim is to perform Bayesian inference about the intensity function $\lambda$ and the parameter vector $\theta$ indexing the model. The full Bayesian model is completely specified by setting a prior distribution $\pi(\theta)$.

The posterior distribution of the unknown quantities of the model ($X$ and $\theta$) is infinite-dimensional and has an intractable density, which makes it unfeasible to devise a straightforward MCMC algorithm to sample from this distribution. We resort to results related to the exact simulation of diffusions \citep[see][]{sermai} to introduce auxiliary variables that allow us to devise a tractable finite-dimensional MCMC algorithm. Those variables define a finite-dimensional representation of the diffusion which can be sampled exactly from its full conditional distribution and such that, conditional on this representation, the parameters indexing the model are independent of the infinite-dimensional remainder of the diffusion and have a tractable full conditional density. Finally, this approach also allows for exact sampling from the posterior distribution of the infinite-dimensional remainder of the diffusion.

\subsection{Theoretical background on diffusions and Poisson processes}\label{back_sec}

We present some theoretical results that are used to develop our methodology. We start with the likelihood function $L(X,\theta)$ of the Poisson process $N$, which is obtained by writing the density of a $PP(\lambda_s)$ w.r.t. to the measure of a $PP(1)$ and is given by
\begin{equation}\label{PP_lik}
\ds L(X,\theta)=\exp\left\{-\int_{0}^{T}(g(X_s;\theta)-1)ds\right\}\prod_{j=1}^{n}g(X_{t_j};\theta).
\end{equation}
Due to the infinite dimensionality of $X$, the likelihood function above is intractable, in the sense that it cannot be analytically computed for arbitrary values of $X$, $\theta$ and $N$.

The density of the diffusion $X$ in $(t_0,t_1]$, for $0\leq t_0<t_1$, conditional on the value of $X_{t_0}$, w.r.t. the measure of a brownian motion with same initial value is obtained using Girsanov's formula and is given by:
\begin{equation}\label{Girsanov1}
\ds \exp\left\{A(X_{t_1};\theta)-A(X_{t_0};\theta)-\int_{t_0}^{t_1}\left(\frac{\alpha^2+\alpha'}{2}\right)(X_s;\theta)ds\right\}.
\end{equation}
where $\ds A(u;\theta)=\int_{0}^u\alpha(y;\theta)dy$.

The density of a diffusion bridge of $X$ in $(t_0,t_1)$, for $0\leq t_0<t_1$, conditional on the values of $X_{t_0}$ and $X_{t_1}$, w.r.t. to the measure of a brownian bridge with same initial and end values is given by the product of (\ref{Girsanov1}) and the term $\ds \frac{f_N(X_{t_1};X_{t_0},t_1-t_0)}{p_{t_1-t_0}(X_{t_0},X_{t_1};\theta)}$, where $f_N(u;a,b)$ is the Lebesgue density of a normal distribution with mean $a$ and variance $b$ and $p_{t_1-t_0}(\cdot,\cdot;\theta)$ is the transition density of the diffusion $X$ in a time interval of size $t_1-t_0$.

We also present a key result involving Poisson processes that shall be useful to derive a rejection sampling algorithm to sample bridges of $X$ from their respective full conditional distributions in the MCMC algorithm to be proposed.
For a fixed path of $X$, consider a function $f(X_s)$ bounded below and above by constants $l$ and $u$, respectively, for $s\in[0,t]$. Let $\Xi$ be a homogeneous Poisson process of intensity $(u-l)$ on $[0,t]\times[0,1]$ and define $N_b$ to be the number of points of $\Xi$ falling below the graph $\{s,(f(X_s)-l)/(u-l);\;s\in[0,t]\}$. Then, standard properties of Poisson process imply that
\begin{equation}\label{Pcoin_exp}
\ds P(N_b=0)=\exp\left\{-\int_{0}^{t}(f(X_s)-l)ds\right\}.
\end{equation}
This result allows us to simulate a Bernoulli random variable with mean given by (\ref{Pcoin_exp}) without the need to compute this value. Let $I$ to be the indicator function of $N_b=0$, then $I$ has the aforementioned Bernoulli distribution and can be simulated by simulating $\Xi$ and unveiling the value of $X$ only and the times instant given by the horizontal coordinates of $\Xi$. We call this the Poisson coin algorithm - formalized in Proposition \ref{rnd3} in Appendix A.

The methodology proposed in this paper can be applied to a wide class $\mathcal{P}$ of models, consisting of those for which we can sample exactly from the distribution of (bridges of) $X$ conditional on the data and parameters. This is done via retrospective rejection sampling where the accept/reject decision is performed by means of the Poisson coin algorithm.
Defining $\mathcal{X}$ as the state space of $X$, the class $\mathcal{P}$ is composed by the DDCP models satisfying the following conditions.
\begin{enumerate}[(a)]
  \item $\alpha$ is differentiable;
  \item $\left(g+{\alpha^2+\alpha'\over 2}\right)(u;\theta)$ is uniformly bounded below by a function of $\theta$, for all $u\in\mathcal{X}$;
  \item $\ds f_N(u;\mu,t)\exp\{-A(u;\theta)\}f_0(u;\theta)$ and $\ds f_N(u;\mu,t)\exp\{A(u;\theta)\}$ are integrable in $u\in\mathcal{X}$, for all $\mu\in\mathds{R}$ and $t>0$;
  \item $g(u;\theta)$ is bounded by $\exp(bu+c)$ for all $u\in\mathcal{X}$, and $(b,c)\in\mathds{R}^2$ - functions of $\theta$.
\end{enumerate}

\subsection{The MCMC algorithm}\label{subsecmcmc}

We propose an MCMC that alternates between updating the diffusion $X$ and the parameters $\theta$ from their respective full conditional distributions. The former is performed via retrospective rejection sampling but is bound to be inefficient if we attempt to update the whole diffusion path in $[0,T]$ at once. That is basically because the acceptance probability of the rejection sampling algorithm decays exponentially as a function of the length of the time interval in which $X$ is to be sampled. We overcome this problem by updating $X$ in sub-intervals of $[0,T]$ defined by a partition of this interval. Note, however, that if this partition is fixed throughout the MCMC, $X$ will never be updated at the times that define the partition and the resulting chain is not irreducible. For that reason, we define the partition of $[0,T]$ to be random and updated at every iteration of the MCMC according to a distribution that is independent of $X$ and $\theta$.

\begin{figure}[!h]\centering
 \includegraphics[width=1\textwidth]{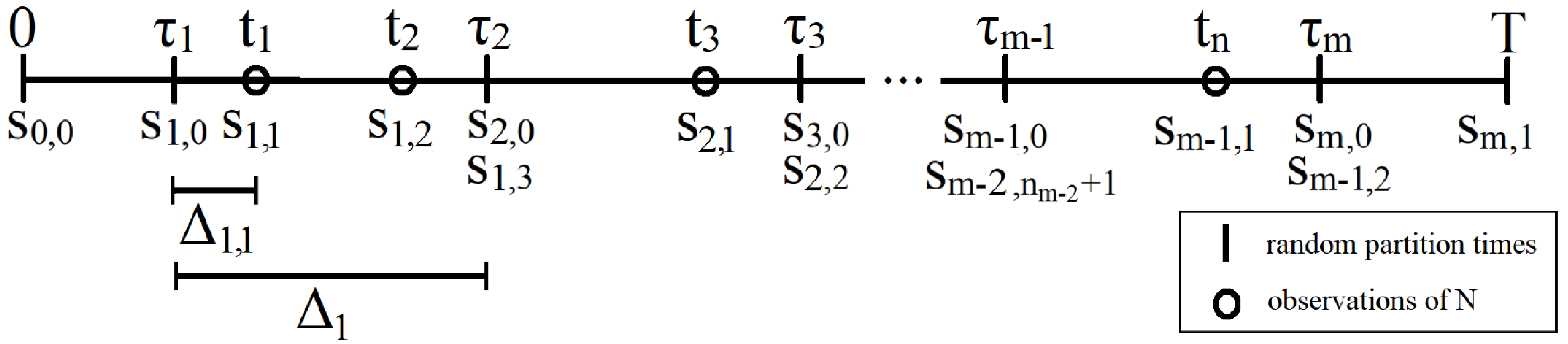}
\caption{An illustration of the notation defined for the time points of interest.}\label{figsketch}
\end{figure}

Let $\tau:=(0=\tau_0<\tau_1<\ldots<\tau_{m}<\tau_{m+1}=T)$ be a partition of the time interval $[0,T]$, with $\Delta_i=\tau_{i+1}-\tau_{i}$. Define $n_i$ to be the number of events from $N$ falling in $[\tau_{i},\tau_{i+1}]$ and let $(s_{i,1},\ldots,s_{i,n_i})$ be those events, with $\Delta_{i,j}=s_{i,j}-s_{i,j-1}$, for $j=1,\ldots,n_{i}+1$, $s_{i,0}=\tau_{i}$ and $s_{i,n_i+1}=\tau_{i+1}$. This means that $s_{i,j}$ is the $j$-th observed event from $N$ in $[\tau_{i},\tau_{i+1}]$. Finally, in order to facilitate the algorithm that samples $X$ in the first and last intervals of the partition, we impose the restriction that $\tau_1$ and $\tau_m$ are such that $n_0=n_{m}=0$. Figure \ref{figsketch} illustrates the notation adopted for the observed events and for the random partition $\tau$ in the interval $[0,T]$.

We now define, for a fixed value of the partition $\tau$, a set of finite-dimensional random variables that, together with $\theta$, constitute the coordinates of the Markov chain in the proposed MCMC algorithm. These random variables are the output of the retrospective rejection sampling algorithm that samples (bridges of) $X$ in each sub-interval defined by the partition $\tau$. This way, the resulting MCMC chain has a finite (albeit random) dimension.

We define $X_{\tau}=(X_0,X_{\tau_1},\ldots,X_{\tau_m},X_T)$ to be $X$ at the partition times and, for $i=1,\ldots,m-1$, $\tilde{X}_i=(X_{s_{i,1}},\ldots,X_{s_{i,n_i}})$ to be $X$ at the event times in $[\tau_{i},\tau_{i+1}]$. For each bridge $(s_{i,j-1},X_{s_{i,j-1}};s_{i,j},X_{s_{i,j}})$ we consider a finite-dimensional measurable function $\mathcal{L}_{i,j}$ of $X$ that defines local upper and lower bounds for $X$ in $[s_{i,j-1},s_{i,j}]$. This function can be simulated, conditional on the start and ending values of the bridge and the value of $\theta$, and is indispensable to perform the Poisson coin algorithm to sample from the full conditional distribution of the bridge. Details about the simulation of $\mathcal{L}_{i,j}$ and how to obtain the aforementioned bounds are provided in Appendix D.

Each set of bounds $\mathcal{L}_{i,j}$ is used to obtain lower and upper bounds on the following function
\begin{equation}\label{phifunc}
\ds \phi(u;\theta)=\left(g+\frac{\alpha^2+\alpha'}{2}\right)(u;\theta).
\end{equation}
For all pair $(i,j)$, let $\phi_{i,j,l}(\theta)$ and $\phi_{i,j,u}(\theta)$ be local lower and upper bounds, respectively, for $\phi(X_s;\theta)$, for $s\in[s_{i,j-1},s_{i,j}]$.

In order to sample $X$ in the intervals defined by the partition $\tau$, with devise a rejection sampling algorithm with proposal distribution given by a biased Brownian motion/bridge. For the first (last) interval in the partition, the initial (end) point of $X$ is also simulated in that interval. For all the other intervals, the simulation is conditional on both the initial and end points.
For each interval $(\tau_{i},\tau_{i+1})$, for $i=1,\ldots,m-1$, the proposal distribution differs from a Brownian bridge $BB(\tau_{i},X_{\tau_{i}};\tau_{i+1},X_{\tau_{i+1}})$ measure only in the distribution of $\tilde{X}_i$. Defining $\pi_{\tilde{\mathbb{W}}_{i}^*}$ and $\pi_{\tilde{\mathbb{W}}_i}$ as the Lebesgue density of $\tilde{X}_i$ under the proposal distribution and under the aforementioned BB measure, respectively, we have that
\begin{equation}\label{densXtj}
\pi_{\tilde{\mathbb{W}}_{i}^*}(\tilde{X}_i;\theta)=\frac{1}{c_i(\theta)}
\pi_{\tilde{\mathbb{W}}_i}(\tilde{X}_i)\times
\prod_{j=1}^{n_i}g(X_{s_{i,j}};\theta),
\end{equation}
where $c_i(\theta)$ is a normalizing constant. Details on how to simulate from the density in (\ref{densXtj}) are provided in Appendix C.

For $i=0$ and $i=m$, the respective intervals contain no events from $N$ and variables $X_0$ and $X_T$ are proposed from the following respective densities.
\begin{eqnarray}
\ds \pi_{\tilde{\mathbb{W}}_{0}^*}(X_0;\theta)&=& \frac{\pi_{\mathbb{W}}(X_0|X_{\tau_1})e^{-A(X_0;\theta)}}{c_0(\theta)}\propto \exp\left\{\frac{-(X_0-X_{\tau_1})^2}{2\tau_1}\right\}f_0(X_0;\theta)e^{-A(X_0;\theta)}, \label{BBM0} \\
\ds \pi_{\tilde{\mathbb{W}}_{m}^*}(X_T;\theta)&=& \frac{\pi_{\mathbb{W}}(X_T|X_{\tau_m})e^{A(X_T;\theta)}}{c_m(\theta)} \propto \exp\left\{\frac{-(X_T-X_{\tau_m})^2}{2(T-\tau_m)}\right\}e^{A(X_T;\theta)}, \label{BBMT}
\end{eqnarray}
where $\mathbb{W}$ is the measure of a Brownian motion with initial distribution $f_0$. The remainder of the proposal is simply a Brownian bridge in $(0,\tau_1)$ and $(\tau_m,T)$, respectively. Simulation from (\ref{BBM0}) and (\ref{BBMT}) may have to be performed indirectly, for example, via rejection sampling. The flexibility to choose $f_0$ is useful to assure the integrability of (\ref{BBM0}).

The acceptance probability of the algorithms that sample $X$ in each interval $(\tau_{i},\tau_{i+1})$, for $i=0,\ldots,m$, of the partition $\tau$ is obtained by combining the expressions in (\ref{PP_lik}), (\ref{Girsanov1}), (\ref{densXtj}), (\ref{BBM0}) and (\ref{BBMT}) and is given by:
\begin{equation}\label{a.p.exp.gen}
\ds \exp\left\{-\int_{\tau_i}^{\tau_{i+1}}\phi(X_s;\theta)-m(\theta)ds\right\},
\end{equation}
where $\ds m(\theta)=\inf_{u\in\mathcal{X}}\{\phi(u;\theta)\}$. Note that this probability has the same form as in (\ref{Pcoin_exp}) and, since a local upper bound for function $\phi$ is available, it can be evaluate by using the Poisson coin algorithm.

We use the local upper and lower bounds $\phi_{i,j,u}(\theta)$ and $\phi_{i,j,l}(\theta)$ so that we can minimize the expected number of time points where $X$ has to be simulated from the proposal distribution. In order to devise a Poisson coin algorithm, we define, for all $(i,j)$,
\begin{equation}\label{r_func}
\ds r_{i,j}(\theta)=\phi_{i,j,u}(\theta)-\phi_{i,j,l}(\theta)
\end{equation}
and a homogeneous Poisson process $\Xi_{i,j}$ with rate $r_{i,j}(\theta)$ on $[s_{i,j-1},s_{i,j}]\times[0,1]$. Let $\kappa_{i,j}$ be the number of events from $\Xi_{i,j}$ and $\Psi_{i,j}=(\psi_{i,1},\ldots,\psi_{i,\kappa_{i,j}})$ and $\Upsilon_{i,j}=(\upsilon_{i,1},\ldots,\upsilon_{i,\kappa_{i,j}})$ be the respective horizontal and vertical coordinates of those events. Now define $\dot{X}_{i,j}=(X_{\psi_{i,1}},\ldots,X_{\psi_{i,\kappa_{i,j}}})$ and set $\ds\tilde{X}:=\{\tilde{X}_i\}_{i=0}^{m}$, $\ds\mathcal{L}_i:=\{\mathcal{L}_{i,j}\}_{j=1}^{n_i+1}$ and $\ds\mathcal{L}:=\{\mathcal{L}_i\}_{i=0}^{m}$, $\ds\dot{X}_i:=\{\dot{X}_{i,j}\}_{j=1}^{n_i+1}$ and $\ds\dot{X}:=\{\dot{X}_i\}_{i=0}^{m}$, $\ds\Xi_i:=\{\Xi_{i,j}\}_{j=1}^{n_i+1}$ and $\ds\Xi:=\{\Xi_i\}_{i=0}^{m}$.

Finally, the acceptance indicator of the aforementioned rejection sampling algorithm, in an interval $[s_{i,j-1},s_{i,j}]$, is given by:
\begin{equation}\label{indrs}
I_i=\mathbb{I}\left[u_i\leq \exp\left\{-\sum_{j=1}^{n_i+1}\Delta_{i,j}(\phi_{i,j,l}(\theta)-m(\theta))\right\}\right] \prod_{j=1}^{n_i+1}\prod_{k=1}^{\kappa_{i,j}}\mathbb{I}\left[\frac{\phi(X_{\psi_{i,j,k}};\theta)-\phi_{i,j,l}(\theta)}{r_{i,j}(\theta)}<\upsilon_{i,j,k}\right],
\end{equation}
where $u_i\sim U(0,1)$. The validity of the algorithm is formally established by Proposition \ref{rnd2} in Appendix A.

Note that $n_0=n_m=0$ by the restriction imposed to the partition $\tau$. If no such restriction was made, the respective proposals would require to be biased not only at the times 0 and $T$, but also at the event times, which could potentially compromise the tractability of the algorithm.

Our MCMC algorithm samples from the posterior of $(\mathcal{L},\tilde{X},\dot{X},\Xi,X_{\tau},\theta)$ in a Gibbs sampling that alternates between sampling $(\mathcal{L},\tilde{X},\dot{X},\Xi,X_0,X_T)$ and $\theta$ from their respective full conditional distributions and the partition $\tau$ from some chosen distribution. An appealing proposal on how to update the partition $\tau$ is presented in Appendix E.

Due to the Markov property of diffusions and Poisson processes, $(\mathcal{L},\tilde{X},\dot{X},\Xi)$ is conditionally independent, given $(X_{\tau},N,\theta)$, among the intervals defined by the partition $\tau$. For each interval $(\tau_{i},\tau_{i+1})$, for $i=1,\ldots,m-1$, $(\mathcal{L}_i,\tilde{X}_i,\dot{X}_i,\Xi_i)$ is sampled using the rejection sampling algorithms described above.

In order to simulate from the full conditional distribution of $\theta$, we obtain the joint density of $(N,\mathcal{L},\tilde{X},\dot{X},\Xi,X_{\tau},\theta)$ with respect to a suitable $\theta$-free dominating measure that guarantees that the full conditional Lebesgue density of $\theta$ is proportional to it. Those two densities are given as follows.

\begin{theorem}\label{maintheo}
For a prior Lebesgue density $\pi(\theta)$ and a fixed value of the partition $\tau$, the joint density of $(N,\mathcal{L},\tilde{X},\dot{X},\Xi,X_{\tau},\theta)$ w.r.t. a $\theta$-free dominating measure is given by
\begin{eqnarray}\label{jpdist}
\ds \pi(N,\mathcal{L},\tilde{X},\dot{X},\Xi,X_{\tau},\theta)&=& \kappa(X_{\tau},\tilde{X})\pi(\theta)f_0(X_0;\theta)\exp\left\{A(X_T;\theta)-A(X_0;\theta)\right\} \nonumber \\
&\times& \exp\left\{\sum_{i=0}^{m}\sum_{j=1}^{n_i+1}\Delta_{i,j}(1-\phi_{i,j,u}(\theta))\right\} \prod_{i=0}^{m}\prod_{j=1}^{n_i}g(X_{s_{i,j}};\theta) \nonumber \\
&\times& \prod_{i=0}^{m}\prod_{j=1}^{n_i+1}\left[r_{i,j}(\theta)^{\kappa_{i,j}}
\prod_{k=1}^{\kappa_{i,j}}\mathbb{I}\left[\frac{\phi(X_{\psi_{i,j,k}};\theta)-\phi_{i,j,l}(\theta)}{r_{i,j}(\theta)}<\upsilon_{i,j,k}\right]\right]
\end{eqnarray}
$\kappa(X_{\tau},\tilde{X})$ is a function of $X_{\tau}$ and $\tilde{X}$ that does not depend on $\theta$ (see proof for details).
\end{theorem}
\begin{thma}
See Appendix F.
\end{thma}
The dominating measure used in (\ref{jpdist}) is fully specified in the proof of the theorem.

By integrating the $\upsilon_{i,j,k}$ variables out in (\ref{jpdist}), we get the following full conditional Lebesgue density of $\theta$.
\begin{eqnarray}\label{fcdt}
\ds \pi(\theta|\cdot) &\propto& \pi(\theta)f_0(X_0;\theta)\exp\left\{A(X_T;\theta)-A(X_0;\theta)-\sum_{i=0}^{m}\sum_{j=1}^{n_i+1}\Delta_{i,j}\phi_{i,j,u}(\theta)\right\} \nonumber \\
&& \prod_{i=0}^{m}\left[\prod_{j=1}^{n_i}g(X_{s_{i,j}};\theta)\right] \left[\prod_{j=1}^{n_i+1}r_{i,j}(\theta)^{\kappa_{i,j}}
\prod_{k=1}^{\kappa_{i,j}}\left(1-\frac{\phi(X_{\psi_{i,j,k}};\theta)-\phi_{i,j,l}(\theta)}{r_{i,j}(\theta)}\right)\right].
\end{eqnarray}
A Metropolis-Hastings step will typically be required to sample from this distribution.

In order to sample from the posterior remainder of $X$, given an MCMC sample from the posterior of $(\mathcal{L},\tilde{X},\dot{X},\Xi,X_{\tau},\theta,\tau)$, we use the following corollary from Theorem \ref{maintheo}.
\begin{corollary}\label{maincorol}
The conditional law of $X$ given $(\mathcal{L},\tilde{X},\dot{X},\Xi,X_{\tau},\theta,\tau,N)$ is independent of $N$ and is given by the joint law of the Brownian bridges between the values of $(X_{\tau},\tilde{X},\dot{X})$, conditional on $\mathcal{L}$.
\end{corollary}
\begin{thma}
See proof of Theorem \ref{maintheo} in Appendix F.
\end{thma}

\subsection{Efficiency of the algorithm}\label{ssecEfAlg}

Note that we are free to choose how to update the partition $\tau$ in the Gibbs sampler. However, this choice has a great impact on the efficiency of the algorithm. In one direction, the smaller the number of sub-intervals is, the lower is the autocorrelation of the chain, leading to faster convergence. On the other hand, the acceptance probability of the rejection sampling algorithm that samples $(\mathcal{L}_i,\tilde{X}_i,\dot{X}_i,\Xi_i)$ decreases (exponentially) as the length of the time interval increases. A reasonable empirical strategy is to choose the minimum number of sub-intervals for which the computational cost is tolerable. Naturally, this depends heavily on functions $\alpha$ and $g$ and on the data.

The computational cost to update $(\mathcal{L}_i,\tilde{X}_i,\dot{X}_i,\Xi_i)$ may substantially vary among the different sub-intervals defined by the partition $\tau$. This is related to the variation of function $\phi$ which, in turn, is related to the information in the data. Typically, time intervals with a higher concentration of observed events will lead to higher variations in the IF and, therefore, higher variations in function $\phi$, resulting in a small acceptance probability. This behavior is usually easy to be identified in each example and the time interval with higher variations can be easily identified in a short pre-run of the MCMC. A reasonable strategy to mitigate the problem is to adopt partitions $\tau$ with different sized intervals. For example, the observed time interval $[0,T]$ is split into sub-intervals of two types such that the partition intervals have different lengths for each type, i.e., at each iteration of the Gibbs sampling, the partition $\tau$ is sampled so that $(\tau_{i+1}-\tau_i)=\varepsilon_j$, if $\tau_i$ is in a sub-interval of type $j$, for $j=1,2$.

The number of time points $m$ defining the partition $\tau$ will be typically large and, therefore, induce a high autocorrelation for the diffusion $X$. This may, in turn, lead to a high autocorrelation of the parameter vector $\theta$. A simple strategy to alleviate this problem is to perform multiple updates of $\tau$ and $(\mathcal{L},\tilde{X},\dot{X},\Xi,X_0,X_T)$ for each update of $\theta$. Furthermore, Monte Carlo estimation should be performed using a thinned sample of $X$.

Finally, note that the algorithm is highly parallelizable due to the conditional independence of the full conditional distributions of diffusion bridges.

Some important parametrization issues related to the efficiency of the MCMC algorithm are discussed in Section \ref{ssecparam}.

\section{Examples}\label{secexamp}

We present some simulated examples to investigate the modeling and inference properties of the proposed methodology. A comparison to an approximate method based on time discretization is presented in Section \ref{secsim2}.

\subsection{Simulated examples with the proposed methodology}\label{secsim1}

We consider two examples for the link function $g$ which have a considerable modeling and inference appeal. The first example is the exponential function, widely used through the well-known Log-Gaussian Cox process \citep{moller,diggle}. The second one is the standard normal cdf $\Phi$, also used for Gaussian process-driven Cox processes \citep{GG}. Those functions feature the nice property of going from $\mathds{R}$ to $\mathds{R}^+$ and $[0,1]$, respectively. We combine those functions with the three diffusion models solving the following SDEs:
\begin{eqnarray}
\ds  dX_s &=& -\rho (X_s-\mu)ds+ dW_s,\;\;\;\rho>0,\;\mu\in\mathds{R}\;\;\;\mbox{(Ornstein-Uhlenbeck - OU)}; \nonumber \\
     dX_s &=& -\rho X_s(\sigma^2X_{s}^2-\mu)ds + dW_s,\;\;\;\rho,\mu,\sigma>0\;\;\;\mbox{(transformed double-well potential - DW)}; \nonumber \\
     dX_s &=& -\frac{X_s}{1+X_{s}^2}ds+dW_s\;\;\;\mbox{(Cauchy)}. \nonumber
\end{eqnarray}
The OU-process is a stationary Gauss-Markov process and the Cauchy process has a Cauchy invariant distribution which allows for longer-term excursions away from 0 than a Gaussian process. The transformed double-well process stochastically alternates visits between two levels (symmetric around 0). In Appendix B, we present some simulated trajectories and Monte Carlo estimates for some models that combine the two link functions above with the OU and the transformed double-well diffusions. The algorithm to simulate from the model is also presented there.

We consider three simulated examples. The first two focus on the analysis of how well the IF and model parameters are estimated and the third one explores model flexibility.
The MCMC chains run for at least 200k iterations with a suitable burn-in and a lag to update the parameters so that the effective sample size of each parameter is at least 500. We use the concept of effective sample size to define the statistics ``time per effective sample" of a parameter as the ratio between the total running time of an algorithm (in seconds) and the effective sample size of that parameter.
We estimate the effective sample size using the R package CODA, which computes the spectral density at frequency zero by fitting an AR model to the chain.
All the estimated parameters in each model are jointly sampled via Metropolis-Hastings with a properly tuned Gaussian random walk proposal.
The algorithms are implemented in Ox \citep{Ox} and run in a 3.50-GHz Intel i7 processor with 6 cores and 16GB RAM.

We fit the exp-OU - $g(X_s)=\exp(\sigma X_s)$, and the cdf-DW - $g(X_s)=\gamma\Phi(\sigma X_s)$, models to one dataset generated from the each of the respective true models in the interval $[0,400]$. The datasets have 499 and 645 events, respectively. The $\sigma$ parameters are not estimated and fixed at their true value. That is because, in those two models, they represent the instant variation of diffusion and are, therefore, weakly identified by the model. Also, note that, for the cdf-DW model, parameter $\rho$ is related to the time spent in each of the two levels and, because there are only a few (roughly 6) level changes in the true IF of the example, this parameter is weakly identified by the data and, therefore, fixed at its true value. The partition $\tau$ is sample at each iteration of the Gibbs sampling with $\tau_{i+1}-\tau_i=1$.

\begin{figure}[!h]\centering
\includegraphics[width=1\textwidth]{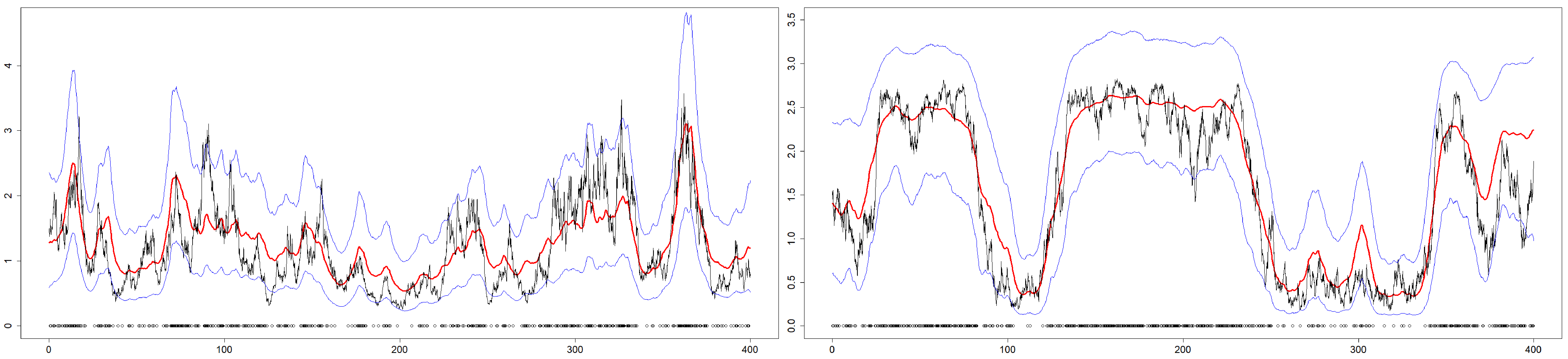}
\caption{Real (black line) and estimated intensity function - posterior mean and pointwise 95\% credibility interval, for the exp-OU (top) and cdf-DW (bottom) examples. The black circles on the bottom represent the data.}\label{figexamp1}
\end{figure}

We adopt $f_0(u)\propto \exp(-\rho\sigma^2u^4/4)$ so that $\ds \pi_{\tilde{\mathbb{W}}_{0}^*}(u)$ is integrable for the cdf-DW example. For the exp-OU example we set $X_0\sim N(\mu,1/(2\rho))$, which is the stationary distribution of the OU-process. Improper uniform priors are adopted for all the parameters. Results are presented in Figure \ref{figexamp1} and Table \ref{tabexamp1}. The posterior correlation of $\mu$ and $\rho$ in the exp-OU model is 0.03 and that of $\gamma$ and $\mu$ in the cdf-DW model is -0.15.

\begin{table}[!h]\centering
\caption{Posterior statistics of the parameters for all the simulated examples.}\label{tabexamp1}
{\scriptsize
\begin{tabular}{|c|ccc|ccc|ccc|ccc|}
  \hline
                        &   \multicolumn{3}{|c|}{$\gamma$} &   \multicolumn{3}{|c|}{$\mu$} & \multicolumn{3}{|c|}{$\rho$} & \multicolumn{3}{|c|}{$\sigma$}   \\ \cline{2-13}
                        & real & mean & sd & real & mean & sd & real & mean & sd & real & mean & sd  \\ \cline{2-13}
                 exp-OU & - & - & - & 0 & 0.14 & 0.13 & 0.05 & 0.095 & 0.030 & 0.2 & - & - \\
                 cdf-DW & 3 & 2.99 & 0.37 & 1 & 1.44 & 0.32 & 0.1 & - & - & 0.2 & - & - \\
             exp-Cauchy & -1.61 & -1.85 & 0.18 & - & - & - & - & - & - & 0.4 & 0.48 & 0.06  \\  \hline
\end{tabular}}
\end{table}

The general class of Cox process models proposed in this paper offers a significant contribution in terms of model flexibility
when compared to the class of models found in the literature, in particular, log-Gaussian Cox processes. In order to illustrate this,
we compare a heavy tailed DDCP - the exp-Cauchy model, to the exp-OU model, which is a type of log-Gaussian Cox process.
The exp-Cauchy model considers the intensity function to be $g(X_s)=\exp(\gamma+\sigma X_s)$, with $X$ being a Cauchy diffusion.
We generate data from this model in $[0,500]$ for $\gamma=log(0.2)\approx-1.61$ and $\sigma=0.4$. The dataset contains 251 points. The partition $\tau$ is sample at each iteration of the Gibbs sampling with $\tau_{i+1}-\tau_i=0.5$. Parameter estimates are presented in Table \ref{tabexamp1}.

We fit the exp-Cauchy and the exp-OU models with fixed $\sigma=0.4$ for both and $f_0(u)\propto e^{-u^2/2}(u^2+1)^{-1/2}$ for the former.
Prediction is performed for two functionals of the intensity function in $[500,1000]$, $\ds I_{\lambda}=\int_{500}^{1000}\lambda_sds$
and $\ds p_{\lambda}=p_4(\lambda_s)$, where $\ds p_4(\lambda_s)$ is the proportion of times the intensity function goes above 4 at time points
multiple of 0.1 in $[500,1000]$.

\begin{figure}[!h]\centering
\includegraphics[width=0.9\textwidth]{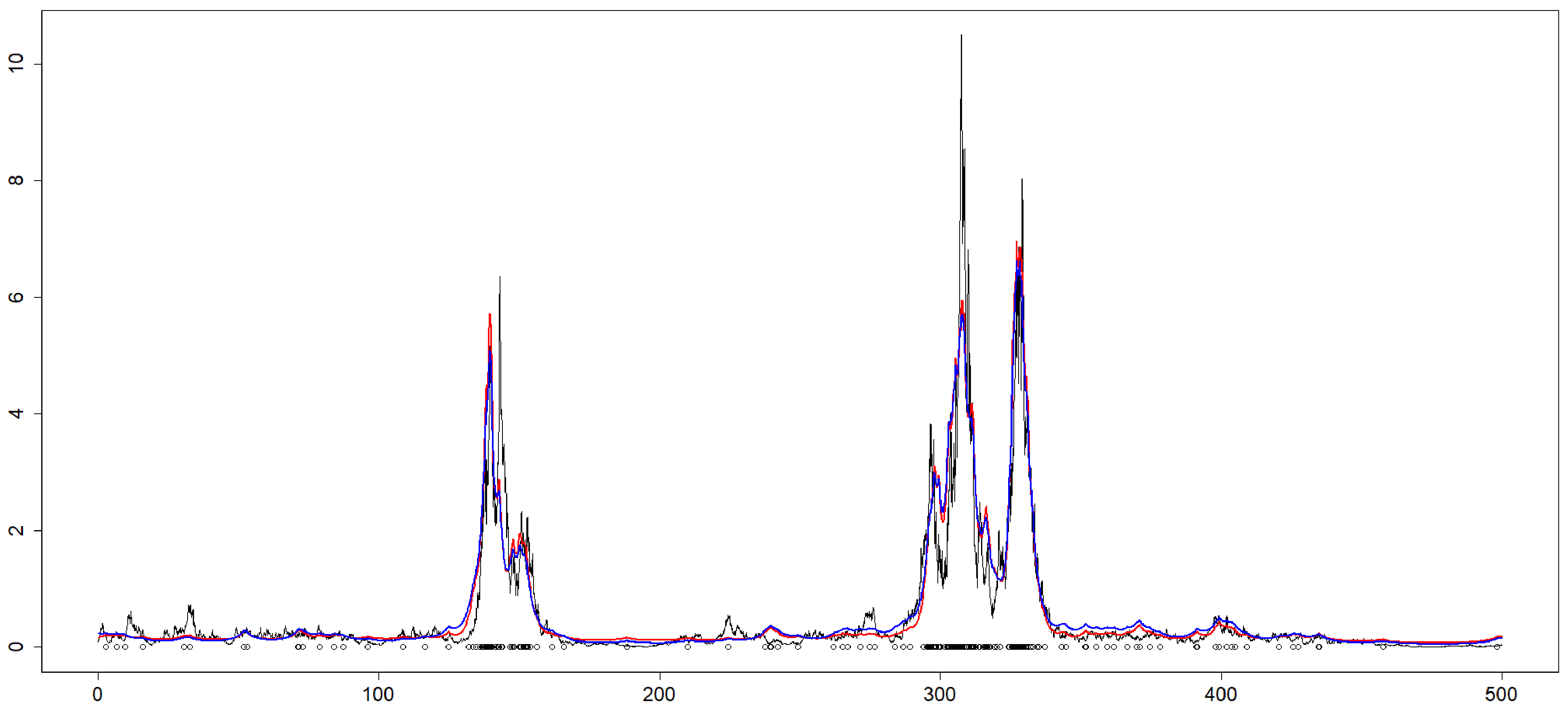}
\caption{Real (black line) and estimated intensity function - under the exp-OU (blue) and exp-Cauchy (red) models. The black circles represent the data.}\label{figexamp4}
\end{figure}

Figure \ref{figexamp4} shows the true IF and the estimated IF under the two models. Table \ref{tabexamp2} show the results for the predictive distribution under the simulation (true) and two fitted models. Although the estimation of the IF in the observed interval is similarly efficient under both models, their predictive power is substantially different. The model misspecification yields highly biased predictions.

\begin{table}[!h]\centering
\caption{Comparison of (percentiles of) the predictive distributions of $I_{\lambda}$ and $p_{\lambda}$ under the estimated exp-OU and exp-Cauchy models and the true (simulation) model.}\label{tabexamp2}
{\scriptsize
\renewcommand{\arraystretch}{1.2}
\begin{tabular}{|c|cccccccccccc|}
  \hline
                        &   \multicolumn{12}{|c|}{$I_{\lambda}$}  \\ \cline{2-13}
                        &  min & 1\%  &  5\% & 10\% &25\% & 50\% & 75\% & 90\% & 95\% & 99\% & 99.9\% & max   \\ \cline{2-13}
                 OU &     2.97  & 38.0 & 69.5 & 89.4 & 131.4 & 198.0 & 318.7 & 554.2 &  859.9 & 2716.9   & 18035.8  & 9.1$\times10^{5}$    \\ \cline{2-13}
                 Cauchy & 0.006 & 46.0 & 67.5 & 77.4 &  95.6 & 126.9 & 225.4 & 940.4 & 4871.4 & 742685.3 & 4.4$\times10^{9}$ & 5.0$\times10^{15}$  \\ \cline{2-13}
                 True & 0.144 & 56.8 & 88.7 & 97.9 &  111.1 & 132.2 & 197.3 & 545.2 & 1851.8 & 87730.0 & 3.4$\times10^{7}$ & 3.9$\times10^{12}$  \\ \cline{2-13}
                 & \multicolumn{12}{|c|}{$p_{\lambda}$} \\ \cline{2-12} \cline{2-13}
                 OU & 0 & 0 & 0 & 0 & 0 & 0 & 0 & 0 & 0 & 0 & 0 & 0\\ \cline{2-13}
                 Cauchy & 0 & 0 & 0 & 0 & 0 & 0.0004 & 0.02 & 0.08 & 0.16 & 0.44 & 0.85 & 0.995 \\ \cline{2-13}
                 True & 0 & 0 & 0 & 0 & 0 & 0 & 0.01 & 0.06 & 0.13 & 0.40 & 0.81 & 0.970  \\ \hline
\end{tabular}}
\end{table}

\subsection{Comparison to a discrete approximation method}\label{secsim2}

We compare the exact methodology proposed in this paper to an approximate one based on time discretization. The latter considers the Euler approximation with time step $\Delta$ for the diffusion $X$ and,
for each interval $(i\Delta,(i+1)\Delta)$, models the number of events observed in that interval as a Poisson distribution with mean $\left(\Delta\times g(X_{i\Delta};\theta)\right)$.

The MCMC algorithm for the discrete model uses the random partition approach proposed in Section \ref{secinf1} to update the diffusion $X$ in each sub-interval via Metropolis Hastings with a Brownian bridge proposal. The parameters are updated via MH with a Gaussian random walk proposal.

We compare the two methodologies for the exp-OU and the exp-Cauchy examples. Detailed results are presented in Appendix G.

As expected, the discrete method has a lower cost to approximate the posterior for the exp-OU model when compared to the exp-Cauchy one, since the OU process is a Gaussian process. Results show a small but non-negligible  difference between the posterior distribution of the parameters of the exp-OU model for the discrete and exact methods. Considering the exp-Cauchy example, results suggest that the discrete approximation is an impracticable option when the true diffusion model is highly non-Gaussian. Although the estimates of the IF were similar between the two methods, the differences regarding the posterior distribution of the parameters is considerable and ought to lead to considerable differences in the predictive distribution.

\subsection{Applications}\label{secapp}

We apply the proposed methodology to three real datasets. The first one is the classic coal mine disaster data of \citet{Jar79}, consisting of the dates of 191 coal mine explosions that killed ten or more men in Britain between March 15th, 1875 and March 22nd, 1962. We consider year as the time unit and the cdf-DW model. Parameters $\delta$ and $\mu$ are estimated and we fix $\rho=0.05$ and $\sigma=0.2$. The second dataset regards the S\&P500 index from Jan 3rd, 2006 to Dec 28th, 2018 - 3270 (working) days, with data consisting of the days in which a variation (w.r.t. the previous day) of more than 20 points occurred - 584 days in total (17.86\%). The exp-OU model is considered with one time unit corresponding to ten days and for fixed $\sigma=0.2$. Finally, the third dataset considers earthquake occurrences in Japan from Jan 1st, 2014 to 30th Dec, 2017, with magnitude 4,5+ (1611 occurrences in 1460 days). The exp-Cauchy model is considered with day as the time unit. The partition $\tau$ is sample at each iteration of the Gibbs sampling with $\tau_{i+1}-\tau_i$. set to be 1 and 0.5 for the first and second examples, respectively. For the third one, we identify time intervals with higher concentration of observed events and apply the strategy described on the second paragraph of Section \ref{ssecEfAlg} with length values $0.5$ and $0.25$, for intervals with lower and higher concentration, respectively.

\begin{figure}[!h]\centering
\centering
\includegraphics[width=1\textwidth]{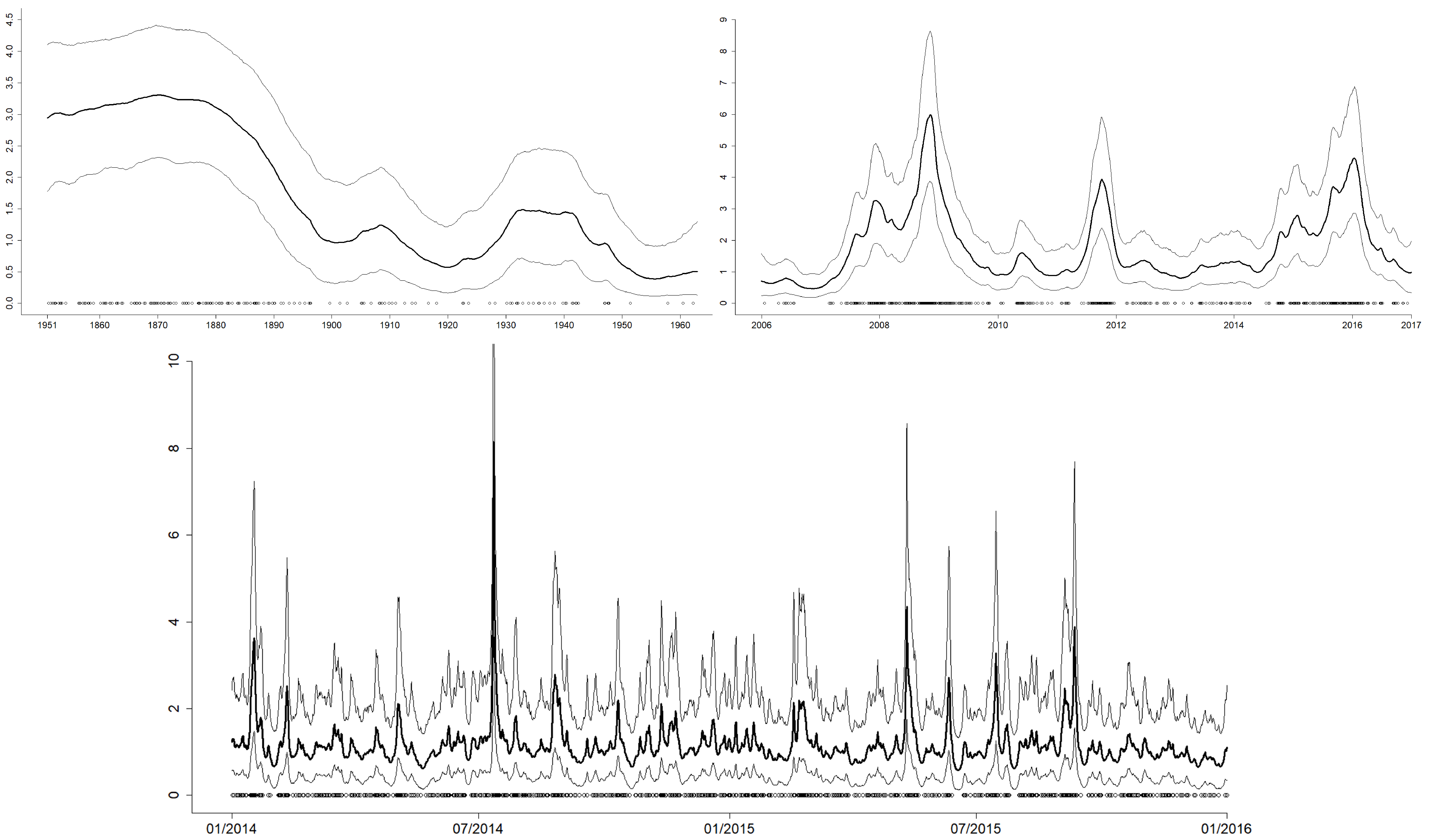}
\caption{Estimated intensity function - posterior mean and pointwise 95\% credibility interval, for the coal mine (top), S\&P500 (middle), and Japan earthquakes (bottom) examples.}\label{figapp0}
\end{figure}

\begin{figure}[!h]\centering
\centering
\includegraphics[width=0.5\textwidth]{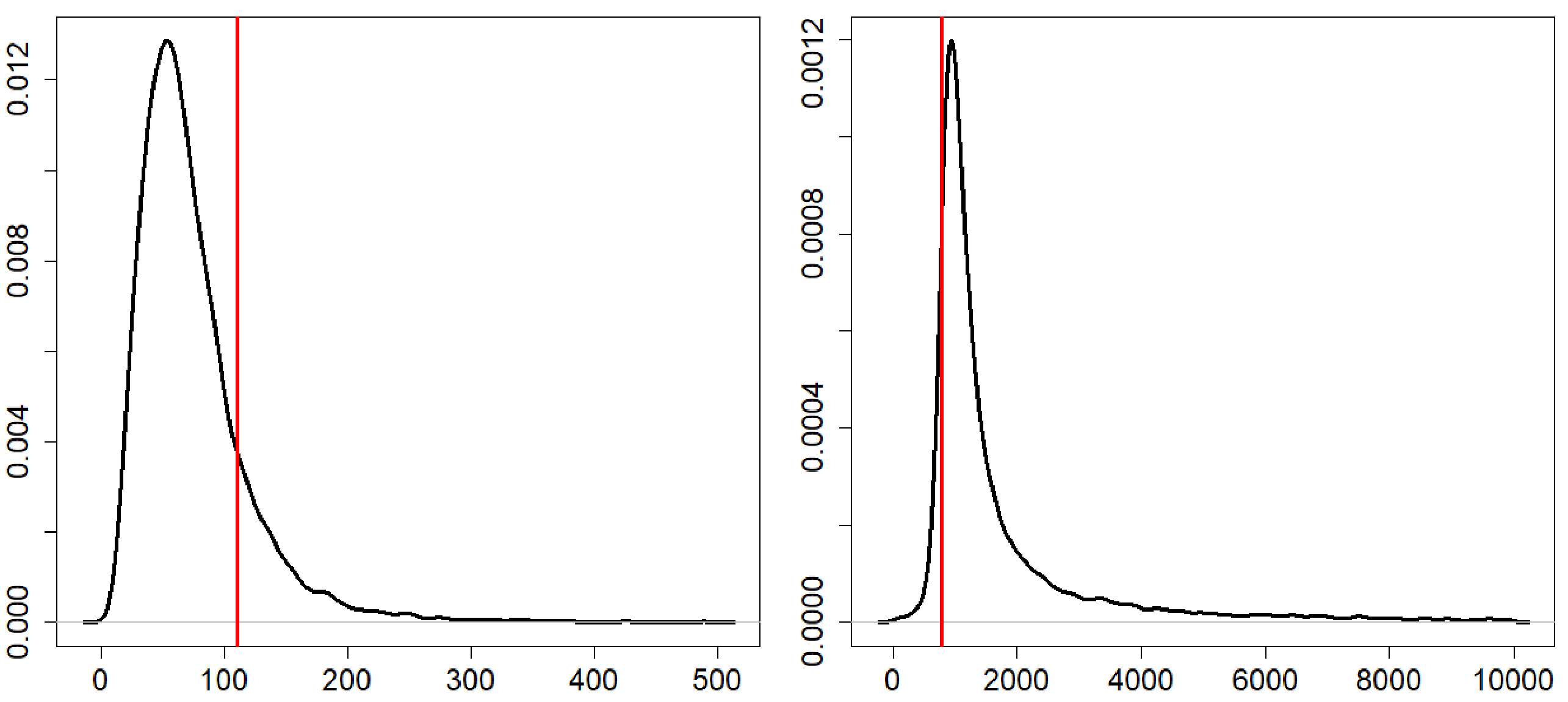}
\caption{Predictive distribution of the integrated IF for the S\&P500 (left) and Japan earthquake (right) examples. The vertical line represents the observed number of events (110 and 778).}\label{figapp2}
\end{figure}

In order to further explore the strengths of the proposed methodology we also present a prediction validation exercise for the S\&P500 and the Japan earthquakes examples. We compare the predictive distribution of the integrated IF in the validation time interval to the true number of events in that interval. We consider the data from 2006 to 2016 (474 events) and predict the next 2 years for the S\&P500 example and the data from 2014 and 2015 (833 events) and predict the next 2 years for the Japan earthquakes one. The predictive distribution for the latter has an extremely heavy right tail so, in order to be able to visualize its density, we plot the empirical density of the truncated (at 10000 - percentile 0.885) predictive distribution.
Results are presented in Figures \ref{figapp0} and \ref{figapp2} and Table \ref{tabapp1}.

\begin{table}[!h]\centering
\caption{Posterior statistics of the parameters for the three applications.}\label{tabapp1}
{\scriptsize
\begin{tabular}{|c|cc|cc|cc|cc|}
  \hline
                        &   \multicolumn{2}{|c|}{$\gamma$} &   \multicolumn{2}{|c|}{$\mu$} & \multicolumn{2}{|c|}{$\rho$} & \multicolumn{2}{|c|}{$\sigma$}   \\ \cline{2-9}
                        & mean & sd & mean & sd & mean & sd & mean & sd  \\ \cline{2-9}
                 cdf-DW & 3.665 & 0.560 & 1.616 & 0.708 &   &   &   & \\
                 exp-OU &  &  & 0.245 & 0.426 & 0.038 & 0.018 &  &  \\
                 exp-Cauchy & 0.019 & 0.061 &  &  &  &  & 0.468 & 0.077 \\  \hline
\end{tabular}
}
\end{table}

\section{Further topics}\label{secFT}

\subsection{A note on parametrization}\label{ssecparam}

The DDCP model defined in (\ref{DDCPeq1})-(\ref{DDCPeq4}) will sometimes admit different parametrizations. Typically, the parametrization of a model has a great impact on the efficiency (convergence properties) of the MCMC algorithm devised to perform inference. We define $\eta_1$ as the set of parameters indexing function $g$ and $\eta_2$ as the set indexing functions $\alpha$ and $f_0$.

We consider the parametrization issue when the same model may be defined with some parameter(s) being either in $\eta_1$ or $\eta_2$, referred to as noncentered and centered parametrizations, respectively. This problem is deeply investigated in \citet{PRSpar}, in a general Gibbs sampling context, who argue that the noncentered parametrization performs better when $X$ (the missing data), under the centered parametrization, is relatively (to the parameter(s) in question) weakly identified by the data. This implies that $X$, under the noncentered parametrization, and the parameter(s) in question are not highly correlated a posteriori, which contributes to the efficiency of the Gibbs sampler.

In the context of DDCP, the diffusion $X$, under the centered parametrization, is strongly identified by the data, relative to the parameters indexing the diffusion, because the data is highly informative about the intensity function. For that reason, the centered parametrization should always be preferred.

It is also important to address the fact that the Poisson process data is, typically, not very informative about the parameters indexing the diffusion. This is basically due to the composition of the Poisson process variance given the IF and the diffusion variance given its parameters. The same phenomenon is observed, for example, for Gaussian process-driven Cox processes \citep[see][]{GG}.

Finally, note that some parameters may eventually appear in both $\eta_1$ and $\eta_2$. That is the case if the model is defined based on a diffusion $Y$ with a diffusion coefficient depending on unknown parameters. Estimation of those parameters is more complicated and, in cases where the data is not very informative about them, these should be fixed at reasonable values, according to the scale of the IF. One way to possibly improve the estimation of those parameters is to devise the MCMC algorithm in terms of $Y=\eta^{-1}(X)$ (the inverse of the Lamperti transform) instead of $X$, so that the centered parametrization can be considered also for parameters in the diffusion coefficient. In practice, the new algorithm will only differ from that in Section \ref{secinf1} in terms of the full conditional distribution of $\theta$. Defining $X_i(\theta)=\eta(Y_{\tau_i};\theta)$ and $X_{i,j}(\theta)=\eta(Y_{s_{i,j}};\theta)$, this distribution is given by
\begin{theorem}\label{theofctheta2}
\begin{eqnarray}\label{fcthetay}
\ds &&\pi(\theta|\cdot)= \pi(\theta)f_{0}^*(Y_0;\theta)\exp\left\{A(X_{m+1}(\theta);\theta)-A(X_0(\theta);\theta)-\sum_{i=0}^{m}\sum_{j=1}^{n_i+1}\Delta_{i,j}\phi_{i,j,u}(\theta)\right\} \nonumber \\
&&\prod_{i=0}^{m} \pi_{\tilde{\mathbb{W}}}(\tilde{X}_i(\theta);X_{i-1}(\theta),X_{i}(\theta)) \frac{1}{\Delta_i}f_{\mathcal{N}}((X_{i}(\theta)-X_{i-1}(\theta))/\Delta_i)\eta'(Y_{\tau_{i}};\theta)\prod_{j=1}^{n_i}\eta'(Y_{s_{i,j}};\theta) \nonumber \\
&&  \prod_{i=0}^{m}\left[\prod_{j=1}^{n_i}g(X_{i,j}(\theta);\theta)\right] \left[\prod_{j=1}^{n_i+1}r_{i,j}(\theta)^{\kappa_{i,j}}
\prod_{k=1}^{\kappa_{i,j}}\left(1-\frac{\phi(\varphi(\omega_{\psi_{i,j,k}};X_{i,j-1}(\theta),X_{i,j}(\theta));\theta)-\phi_{i,j,l}(\theta)}{r_{i,j}(\theta)}\right)\right]. \nonumber
\end{eqnarray}
where $$\ds \omega_{\psi_{i,j,k}}=X_{\psi_{i,j,k}}-\left( 1-\frac{\psi_{i,j,k}-s_{i,j-1}}{s_{i,j}-s_{i,j-1}} \right)X_{i,j-1}(\theta_0) - \left(\frac{\psi_{i,j,k}-s_{i,j-1}}{s_{i,j}-s_{i,j-1}} \right)X_{i,j}(\theta_0),$$
$\theta_0$ is the current value of $\theta$ in the chain used to simulate $X_{\psi_{i,j,k}}$ and
$$\ds \varphi(\omega_{\psi_{i,j,k}};X_{i,j-1}(\theta),X_{i,j}(\theta))=\omega_{\psi_{i,j,k}}+\left( 1-\frac{\psi_{i,j,k}-s_{i,j-1}}{s_{i,j}-s_{i,j-1}} \right)X_{i,j-1}(\theta) + \left(\frac{\psi_{i,j,k}-s_{i,j-1}}{s_{i,j}-s_{i,j-1}} \right)X_{i,j}(\theta).$$
\end{theorem}
\begin{thma}
See Appendix F.
\end{thma}

Another possible reparametrization regards the Poisson processes $\Xi_{i,j}$. Basically, some models may induce a high correlation between the parameters indexing the diffusion and the $\Xi_{i,j}$'s. \citet{sermai} proposes a noncentered parametrization to reduce that dependence that, instead of simulating a Poisson process with rate $r_{i,j}(\theta)$ on $[0,t]\times[0,1]$, simulates a Poisson process with rate 1 on $[0,t]\times[0,\infty)$. This strategy works because we only need to unveil the points for which the second coordinate falls below $r_{i,j}(\theta)$. This leads to the following full conditional distribution for $\theta$:
\begin{theorem}\label{theofctheta3}
\begin{eqnarray}\label{fctheta72}
\ds \pi(\theta|\cdot)&\propto&
\prod_{i=0}^{m}\left[\prod_{j=1}^{n_i}g(X_{i,j};\theta)\right] f_0(X_0;\theta)\pi(\theta) \nonumber \\
&&\exp\left\{A(X_T;\theta)-A(X_0;\theta)-\sum_{i=0}^{m}\sum_{j=1}^{n_i+1}\Delta_{i,j}\phi_{i,j,l}(\theta)\right\} \nonumber \\
&&\prod_{i=0}^{m}\prod_{j=1}^{n_i+1}\left[\prod_{k=1}^{\infty}\left(1-\mathbb{I}(\upsilon_{i,j,k}<r_{i,j}(\theta))\frac{\phi(X_{\psi_{i,j,k}};\theta)-\phi_{i,j,l}(\theta)}{r_{i,j}(\theta)}\right)\right].
\end{eqnarray}
\end{theorem}
\begin{thma}
Combines the proof of our Theorem \ref{maintheo} and that of Theorem 3 from \citet{sermai}.
\end{thma}
The practical computational difference between this algorithm and the original one is that, when proposing a move from $\theta$ to $\theta^*$, one needs to simulate potential extra points from $\Xi_{i,j}$ if $r_{i,j}(\theta^*)>r_{i,j}(\theta)$ \citep[see][Section 4.1]{sermai}. The double-well potential process is a typical example in which the noncentered parametrization leads to significant improvement, so this was applied to the cdf-DW examples presented in Sections \ref{secsim1} and \ref{secapp}.

\subsection{Prediction}\label{secinf4}

Prediction about future behavior is generally of interest when fitting unidimensional Cox processes. Under the Bayesian Paradigm, that is naturally achieved through the predictive distribution, i.e., the posterior distribution of some function of the process in some unobserved time interval.

It is straightforward to obtain a sample from the predictive distribution in an MCMC context, as it is considered in this paper. Suppose that we want to predict some function $h(N^*,X^*)$ of the Cox process and the diffusion $X$ in some unobserved time interval, given the data $N$. The predictive distribution of $h(N^*,X^*)$ satisfies:
\begin{equation}\label{predct}
\ds \pi(h(N^*,X^*)|N)=\int\pi(h(N^*,X^*)|X,\theta)\pi(X,\theta|N)dXd\theta.
\end{equation}
This means that a sample from the predictive distribution $\pi(h(N^*,X^*)|N)$ can be obtained by simulating one observation from $\pi(h(N^*,X^*)|X^{(j)},\theta^{(j)})$, for each $(X^{(j)},\theta^{(j)})$ in the posterior sample output in the MCMC algorithm. Naturally, simulation from $\pi(h(N^*,X^*)|X,\theta)$ should be possible.

Consider for example $h(N^*,X^*)=N_{T+t^*}-N_T$, for some $t^*>0$, i.e. the number of events in the interval of length $t^*$ following the observed interval. Simulation from $\pi(h(N^*,X^*)|N)$ is performed as follows:
\\{\scriptsize
\begin{tabular}[!]{|l|}
\hline\\
\parbox[!]{13cm}{
\texttt{
{\bf Simulation from the predictive distribution}
\begin{enumerate}
\setlength\itemsep{-0.5em}
\item initiate $j=1$;
\item simulate $X^{*(j)}:=\{X_t;T\leq t \leq T+t^*\}$, given $X_{T}^{(j)}$ and $\theta^{(j)}$, including $\mathcal{L}$;
\item obtain an upper bound $\lambda^{*(j)}$ for $\lambda_s$ in $[T,T+t^*]$ using $\mathcal{L}$;
\item simulate a Poisson process with rate $\ds \lambda^{*(j)}$ on $[0,t^*]$;
\item simulate $X^{*(j)}$ at the times of the $PP(\lambda^{*(j)})$;
\item keep each of the events $t_i$ with probability $\ds R(t_i)=\frac{g(X_{t_i}^{*(j)};\theta^{(j)})}{\lambda^{*(j)}}$.
\item store $N_{T+t^*}^{(j)}-N_T$, make $j=j+1$ and go to 2 until the whole MCMC sample is used.
\end{enumerate}
}}\\  \hline
\end{tabular}}\\
Steps 2 and 5 are performed via the EA algorithm \citep{bpr07} that performs exact simulation of a class of diffusion processes.

It is also feasible to devise unbiased Monte Carlo estimators for expectations of intractable functions $h$ under the predictive distribution. For example, suppose that we want to estimate $I=E_{(X^*,\theta|N)}\left[\int_{T}^{T+t^*}g(X_s;\theta)ds\right]$. Unbiased estimation is achieved by defining a r.v. $U$ with uniform distribution in $(T,T+t^*)$ and noting that $I=t^*E_{(U,X_0,\theta|N)}\left[g(X_U;\theta)\right]$. This means that an unbiased estimator of $I$ is given by $\hat{I}=t^*\frac{1}{J}\sum_{j=1}^Jg(X_{U^{(j)}}^{(j)};\theta^{(j)})$, which can be computed by simulating $J$ iid samples of $U$ and $X_{U}$ (the latter from its predictive distribution).

\subsection{Inference for different data schemes and extensions}\label{IDDS}

\subsubsection{Incomplete interval}

Suppose that the Poisson process $N$ is observed in two disjoint intervals $[0,T_1]$ and $[T_2,T]$, for $T_1<T_2$ and we want to estimate the intensity function in those intervals and also in the unobserved interval $[T_1,T_2]$. The methodology proposed in Section 2 of the paper can be adapted to perform this task as follows.

The whole interval $[0,T]$ is partitioned by $\tau$. For intervals $(\tau_{i},\tau_{i+1})$ with $\tau_{i}>T_2$ or $\tau_{i+1}<T_1$ $X$ is sampled using the same algorithm from Section 2.2 of the paper.

Intervals $(\tau_{i},\tau_{i+1})$ that contain either $T_1$ or $T_2$ are also sampled via rejection sampling by proposing from the biased Brownian bridge defined in Section 2.2 of the paper and with acceptance indicator given by:
\begin{eqnarray}\label{indrs2}
I_i&=&\mathbb{I}\left[\exp\left(-\sum_{j=1}^{n_i+1}\Delta_{i,j}(\phi_{i,j,l}(\theta)-m(\theta))\right)<u\right] \prod_{j=1}^{n_i+1}\prod_{k=1}^{\kappa_{i,j}}\mathbb{I}\left[\frac{\phi(X_{\psi_{i,j,k}};\theta)-\phi_{i,j,l}(\theta)}{r_{i,j}(\theta)}<\upsilon_{i,j,k}\right] \nonumber \\
&\times& \mathbb{I}\left[\exp\left(-\Delta(\ddot{\phi}_{i,l}(\theta)-\ddot{m}(\theta))\right)<\ddot{u}\right] \prod_{k=1}^{\ddot{\kappa}_{i}}\mathbb{I}\left[\frac{\ddot{\phi}(X_{\ddot{\psi}_{i,k}};\theta)-\ddot{\phi}_{i,l}(\theta)}{\ddot{r}_{i}(\theta)}<\ddot{\upsilon}_{i,k}\right],
\end{eqnarray}
where $u$ and $\ddot{u}$ are independent uniform $(0,1)$, $\Delta=(\tau_{i+1}-T_1)$, if $T_1\in(\tau_{i},\tau_{i+1})$, and $\Delta=(T_2-\tau_{i})$, if $T_2\in(\tau_{i},\tau_{i+1})$. Also, $\ds \ddot{\phi}(u;\theta)=\left(\frac{\alpha^2+\alpha'}{2}\right)(u;\theta)$ and the notation with two dots on top are defined in terms of $\ddot{\phi}$ as the original notation is defined in terms of $\phi$. Components on the first row of (\ref{indrs2}) correspond to interval $(\tau_{i},T_1)$, if $T_1\in(\tau_{i},\tau_{i+1})$, and to interval $(T_2,\tau_{i+1})$, if $T_2\in(\tau_{i},\tau_{i+1})$, and components on the second row correspond to interval $(T_1,\tau_{i+1})$, if $T_1\in(\tau_{i},\tau_{i+1})$, and to interval $(\tau_{i},T_2)$, if $T_2\in(\tau_{i},\tau_{i+1})$.

Intervals $(\tau_{i},\tau_{i+1})$ for which $\tau_{i}>T_1$ and $\tau_{i+1}<T_2$ are sampled via rejection sampling by proposing from a Brownian bridge and with acceptance indicator given by:
\begin{equation}\label{indrs3}
I_i=\mathbb{I}\left[\exp\left(-(\tau_{i+1}-\tau_{i})(\ddot{\phi}_{i,l}(\theta)-\ddot{m}(\theta))\right)<\ddot{u}\right] \prod_{k=1}^{\ddot{\kappa}_{i}}\mathbb{I}\left[\frac{\ddot{\phi}(\dot{X}_{\ddot{\psi}_{i,k}};\theta)-\ddot{\phi}_{i,l}(\theta)}{\ddot{r}_{i,j}(\theta)}<\ddot{\upsilon}_{i,k}\right].
\end{equation}

Finally, the full conditional density of the parameter vector $\theta$ is proportional to
\begin{eqnarray}\label{jpdist2}
\ds \pi(N,\mathcal{L},\tilde{X},\dot{X},\Xi,X_{\tau},\theta)&=& \kappa(X_{\tau},\tilde{X})\pi(\theta)f_0(X_0;\theta)\exp\left\{A(X_T;\theta)-A(X_0;\theta)\right\} \\
&\times& \exp\left\{\sum_{i}\sum_{j=1}^{n_i+1}\Delta_{i,j}(1-\phi_{i,j,u}(\theta))\right\} \prod_{i}\prod_{j=1}^{n_i}g(X_{s_{i,j}};\theta) \nonumber \\
&\times& \prod_{i}\prod_{j=1}^{n_i+1}\left[r_{i,j}(\theta)^{\kappa_{i,j}}
\prod_{k=1}^{\kappa_{i,j}}\mathbb{I}\left[\frac{\phi(X_{\psi_{i,j,k}};\theta)-\phi_{i,j,l}(\theta)}{r_{i,j}(\theta)}<\upsilon_{i,j,k}\right]\right], \nonumber \\
&\times& \exp\left\{\sum_{i}(\tau_{i+1}-\tau_{i})(1-\ddot{\phi}_{i,u}(\theta))\right\} \nonumber \\
&\times& \prod_{i}\left[\ddot{r}_{i}(\theta)^{\ddot{\kappa}_{i}}
\prod_{k=1}^{\ddot{\kappa}_{i}}\mathbb{I}\left[\frac{\ddot{\phi}(X_{\ddot{\psi}_{i,k}};\theta)-\ddot{\phi}_{i,l}(\theta)}{\ddot{r}_{i}(\theta)}<\ddot{\upsilon}_{i,k}\right]\right], \nonumber
\end{eqnarray}

Note that the posterior predictive distribution of the IF in $[T_1,T_2]$ is already sampled from in the MCMC algorithm. This way, prediction for the point process in $[T_1,T_2]$ is performed by sampling from a Poisson process with the IF sampled from the aforementioned predictive distribution.

\subsubsection{Aggregated data}

Suppose that instead of observing the complete Poisson process $N$, we only observe the number of points in a collection of intervals defining a partition of $[0,T]$. This is a common feature in real datasets in which data are aggregated in small time intervals (w.r.t. the total observed time interval), like daily counting data for processes observed over weeks/months.

Define $\Gamma:=(0=\gamma_0,\gamma_1,\ldots,\gamma_p=T)$ to be the $p+1$ time points defining the partition in which the aggregated data is observed and $\dot{n}:=(\dot{n}_1,\ldots,\dot{n}_p)$ to be the number of events in each of the intervals from the partition. Poisson process properties imply that all the $\dot{n}_j$'s are conditionally independent with $\ds\dot{n}_j\sim Poisson\left(\int_{\gamma_{j-1}}^{\gamma_j}g(X_s;\theta)ds\right)$.
This leads to the likelihood function $\ds L(X,\theta)\propto\exp\left\{-\int_{0}^{T}g(X_s;\theta)ds\right\} \prod_{j=1}^{p}\left(\int_{\gamma_{j-1}}^{\gamma_j}g(X_s;\theta)ds\right)^{\dot{n}_j}$.

Inference for this data scheme can be carried out for models in the $\mathcal{P}$ class that, additionally, have a bounded intensity function conditional on the parameters, for example, the cdf-$^{\ast}$ models. Suppose that $M(\theta)$ is an upper bound for $g(u;\theta),\;u\in\mathcal{X}$. We consider a Gibbs sampling algorithm analogous to the one proposed in Section 2.2 of the paper. The partition $\tau$ of $[0,T]$ needs to be discretely distributed over $\Gamma$ such that each interval $(\tau_{i},\tau_{i+1})$ needs to contain at least two of the subintervals defined by $\Gamma$, so that $\tau$ can be updated to assure irreducibility of the chain.

The diffusion bridges are sampled in each subinterval $(\tau_i,\tau_{i+1})$ via rejection sampling by proposing from a Brownian bridge and accepting with probability:
\begin{equation}\label{DDS3}
\ds \alpha_1=\exp\left\{-\int_{\tau_i}^{\tau_{i+1}}\phi(X_s;\theta)-m(\theta)ds\right\} \prod_{j\in\tau_{i,i+1}}\left(\frac{\int_{\gamma_{j-1}}^{\gamma_j}g(X_s;\theta)ds}{(\gamma_j-\gamma_{j-1})M(\theta)}\right)^{\dot{n}_j},
\end{equation}
for $\phi$ and $m(\theta)$ as previously defined and $\tau_{i,i+1}$ representing the set of observed intervals that define $(\tau_i,\tau_{i+1})$. The acceptance probability in (\ref{DDS3}) is evaluated by applying the Poisson coin algorithm for the exponential term and by computing an unbiased (and a.s. in $(0,1)$) estimator for the product term.

Define $r_{i}$ and $\phi_{i,l}$ as in Section 2.2 of the paper, by making $n_i=0$, and consider the indicator function
\begin{equation}\label{indrs4}
I_i=\mathbb{I}\left[\exp\left(-\Delta_{i}(\phi_{i,l}(\theta)-m(\theta))\right)<u\right] \prod_{k=1}^{\kappa_{i}}\mathbb{I}\left[\frac{\phi(\dot{X}_{\psi_{i,k}};\theta)-\phi_{i,l}(\theta)}{r_{i}(\theta)}<\upsilon_{i,k}\right],
\end{equation}
where $u\sim U(0,1)$.

Furthermore, an unbiased estimator of the product term in (\ref{DDS3}) is given by
\begin{equation}\label{DDS4}
\ds \prod_{j\in\tau_{i,i+1}}\prod_{k=1}^{\dot{n}_j}\frac{g(X_{U_{j,k}};\theta)}{M(\theta)},
\end{equation}
where $U_j=(U_{j,1},\ldots,U_{j,\dot{n}_j})$ and the $U_{j,k}$'s are i.i.d. $U(\gamma_{j-1},\gamma_j)$. Finally, the initial and end intervals are sampled by proposing from a biased Brownian motion which biases the extreme points with terms $\exp\{-A(X_0;\theta)\}$ and $\exp\{A(X_T;\theta)\}$, respectively. A proposal bridge is then accepted if $I_i=1$ and a simulated Bernoulli r.v. with success probability given by (\ref{DDS4}) returns 1.

The full conditional density of the parameter vector $\theta$ is derived analogously to Theorem 1 and is given by:
\begin{eqnarray}\label{DDS5}
\ds \pi(\theta|\cdot)&\propto&
\left[M(\theta)\right]^{\ds\sum_{j=1}^p\dot{n}_j} f_0(X_0;\theta)\pi(\theta) \nonumber \exp\left\{A(X_T;\theta)-A(X_0;\theta)-
\sum_{i=0}^{m}\Delta_i\phi_{i,u}(\theta)\right\} \nonumber \\
&&\prod_{i=0}^{m}\left[\left(r_{i}(\theta)\right)^{\kappa_{i}}
\prod_{k=1}^{\kappa_{i}}\left(1-\frac{\phi(\dot{X}_{\psi_{i,k}};\theta)-\phi_{i,l}(\theta)}{r_{i}(\theta)}\right)\right].
\end{eqnarray}

\subsubsection{Extensions}

\citet{GRL} devise a general MCMC algorithm to perform exact inference for discretely observed (jump-)diffusion processes. The algorithm consists of a Gibbs sampling that alternates between updating parameters and missing paths between observations. Each of those two blocks is updated via Barker's steps in which the unknown acceptance probabilities are evaluated using a Bernoulli Factory \citep[see][]{p2p} called the Two-Coin algorithm. This could be adapted to DDCPs by incorporating the Poisson process likelihood (\ref{IRD1}) to the expression of the acceptance probability of seach of the Barker's steps so to extended the proposed methodology to consider DDCPs outside the class $\mathcal{P}$, requiring the drift $\alpha$ to be differentiable.

\section{Conclusions}\label{secconcl}

This paper proposes the first exact methodology to perform inference in a class of diffusion-driven Cox processes. The methodology is exact in the sense that no discretization-based approximation is used and MCMC error is the only source of inaccuracy. The proposed MCMC algorithm is a Gibbs Sampling that alternates between updating the diffusion path and the parameters indexing the model.

The exactness feature of the algorithm lies in the key fact that the global acceptance probability of the rejection sampling algorithm that samples from the full conditional distribution of the diffusion (bridges) has an intractable term which also appears in the joint density of the data and the diffusion at a finite collection of time points.

Several issues related to model flexibility and the efficiency of the proposed methodology are discussed and illustrated in simulated examples.
Results show a very good recovery of the intensity function and the Cauchy diffusion example illustrates the model flexibility when compared to the most popular models in the literature. Comparisons to a discretization-based method show the advantages of the exact methodology. Finally, three real data examples are presented, concerning coal mine accidents, the SP500 index and earthquakes in Japan.

Further discussions are presented regarding model parametrization, prediction and inference for the cases where the data is aggregated or not completely observed. The centered parametrization is argued to have a better performance. Prediction is straightforwardly performed with an extra sampling step that uses the MCMC output to sample from the desired predictive distribution. Finally, an extension of the proposed methodology for a wider class of models is discussed with the use of an infinite-dimensional Barker's MCMC algorithm.

\section*{Computer codes and data}

The computer codes and data that supports the findings of this study are openly available in GitHub at https://github.com/fbambirra/DDCP.git.

\section*{Acknowledgements}

We would particularly like to thank the two anonymous referees who provided excellent and detailed comments on earlier versions of this paper.
Fl\'{a}vio Gon\c{c}alves
would like to thank FAPEMIG - grants PPM-00745-18 and APQ-01837-22, CNPq - grant 310433/2020-7 and the University of Warwick, for financial support. Krzysztof {\L}atuszy\'nski is supported by the Royal Society through the  Royal Society University Research Fellowship.  Gareth Roberts is supported by the EPSRC grants: {\em ilike}
(EP/K014463/1), CoSInES (EP/R034710/1) and Bayes for Health (EP/R018561/1).

\newpage

\section*{Appendix A - Important results and definitions}

Let $\tilde{\mathbb{P}}$ be the probability law of the diffusion in (\ref{DDCPeq3})-(\ref{DDCPeq4}) and $\tilde{\mathbb{P}}_i$, for $i=1,\ldots,m-1$, be the probability measure of $\ds(X^{(i)}|X_{\tau_{i}},X_{\tau_{i+1}},N,\theta,X^{(-i)})$, where $X^{(i)}$ is $X$ in $(\tau_{i},\tau_{i+1})$ and $X^{(-i)}$ is $X$ elsewhere. Let also $\tilde{\mathbb{P}}_0$ be the probability measure of $\ds(X_0,X^{(0)}|N,X_{\tau_1},\theta,X^{(-0)})$, where $X^{(0)}$ is $X$ in $(0,\tau_1)$ and $X^{(-0)}$ is $X$ elsewhere, and $\tilde{\mathbb{P}}_{m}$ be the probability measure of $\ds(X_T,X^{(m)}|N,X_{\tau_m},\theta,X^{(-m)})$, where $X^{(m)}$ is $X$ in $(\tau_m,T)$ and and $X^{(-m)}$ is $X$ elsewhere. Now define $\tilde{\mathbb{P}}$ to be the probability measure of $\ds(X|X_{\tau_{1}},\ldots,X_{\tau_{m}},N,\theta)$.

Let $\mathbb{W}$ be a Brownian motion in $[0,T]$ with initial distribution $f_0$, $\tilde{\mathbb{W}}_i$ be the Brownian bridge $BB(\tau_{i},X_{\tau_{i}};\tau_{i+1},X_{\tau_{i+1}})$, for $i=1,\ldots,m-1$, $\ds \mathbb{W}_{0}$ be the measure of a Brownian motion in $[0,\tau_1)$ with initial distribution $f_0$ and $\ds \mathbb{W}_{m+1}$ be the measure of a Brownian motion in $(\tau_m,T]$. Let also $\tilde{\mathbb{W}}_{i}^*$, $\mathbb{W}_{0}^*$ and $\mathbb{W}_{m+1}^*$ be the measures of the respective biased Brownian bridges/motions defined in (\ref{densXtj}), (\ref{BBM0}) and (\ref{BBMT}), and define the product measure $\tilde{\mathbb{W}}=\mathbb{W}_0\otimes\tilde{\mathbb{W}}_i\otimes\ldots\otimes\tilde{\mathbb{W}}_m\otimes\mathbb{W}_{m+1}$.

Finally, let $\Xi^+$ be the product measure of $n+m+1$ unit rate Poisson processes on $[s_{i,j-1},s_{i,j}]\times[0,1]$, $\forall\;i,j$, $\mathbb{L}$ be the Lebesgue measure on $\mathds{R}$ and $\mathbb{N}$ be the measure of a unit rate Poisson process on $[0,T]$. We also define the extension of the measures $\tilde{\mathbb{W}}$ and $\tilde{\mathbb{W}}^*$ to the space of $N$ to be a unit rate Poisson process on $[0,T]$.

We set, for $i=0,\ldots,m$,
\begin{eqnarray}\label{def_eqs}
\ds L(X,\theta)&\propto&\exp\left\{-\int_{0}^{T}g(X_s;\theta)ds\right\}\prod_{j=1}^{n}g(X_{t_j};\theta), \label{IRD1}\\
\ds L_i(X,\theta)&\propto&\exp\left\{-\int_{\tau_{i}}^{\tau_{i+1}}g(X_s;\theta)ds\right\}\prod_{j=1}^{n_i}g(X_{s_{i,j}};\theta), \label{IRD2}\\
\ds \mathcal{G}(X,\theta) &=& \exp\left\{-\int_{0}^{T}\left(\frac{\alpha^2+\alpha'}{2}\right)(X_s;\theta)ds\right\}, \label{IRD3}\\
\ds \mathcal{G}_i(X,\theta) &=& \exp\left\{-\int_{\tau_{i}}^{\tau_{i+1}}\left(\frac{\alpha^2+\alpha'}{2}\right)(X_s;\theta)ds\right\}. \label{IRD4}
\end{eqnarray}

The proofs of all the results below are presented in Appendix F.

\begin{proposition}\label{rnd1}
Define $X_{-\tau}:=X\setminus X_{\tau}$. Then
\begin{equation}\label{fcX2}
\ds \frac{d\tilde{\mathbb{P}}}{d\tilde{\mathbb{W}}}(X_{-\tau})=e^{T}L(X,\theta)\mathcal{G}(X,\theta)\exp\left\{A(X_T;\theta)-A(X_0;\theta)\right\}\frac{d\mathbb{W}}{d\mathbb{P}}(X_{\tau},N|\theta).
\end{equation}
\end{proposition}

\begin{proposition}\label{POD}
\begin{eqnarray}\label{fcext1}
\ds \frac{d\tilde{\mathbb{W}}}{d\tilde{\mathbb{W}}^*}(X_{-\tau}) = c(\theta) \exp\left\{A(X_0;\theta)-A(X_T;\theta)\right\} \frac{1}{\prod_{j=1}^{n}g(X_{t_j};\theta)},
\end{eqnarray}
where $c(\theta)=\prod_{i=0}^{m}c_i(\theta)$.
\end{proposition}

\begin{proposition}\label{rnd2}
The acceptance probability of the rejection sampling algorithms for the diffusion paths in $[0,\tau_1)$, $[\tau_{i},\tau_{i+1})$, for $i=1,\ldots,m-1$, and $[\tau_m,T)$, described in Section \ref{subsecmcmc}, is given by
\begin{equation}\label{fcX2}
\ds \exp\left( -\int_{\tau_{i}}^{\tau_{i+1}}\left(\phi(X_s;\theta)-m(\theta)\right)ds \right). \nonumber
\end{equation}
\end{proposition}

\begin{proposition}\label{rnd3}
Consider function $\phi(\cdot;\theta)$ as defined in Section \ref{secinf1}. Now let $r_{i,j}(\theta)$ be an upper bound for the function $\phi(X_s;\theta)-\phi_{i,j,l}(\theta)$ in $[s_{i,j-1},s_{i,j}]$ and $\Xi_{i,j}$ be a homogeneous Poisson process of intensity $r_{i,j}(\theta)$ on $[0,t]\times[0,1]$. Now define $N_b$ to be the number of points of $\Xi_{i,j}$ falling below the graph $\{(s,(\phi(X_s;\theta)-\phi_{i,j,l}(\theta))/r_{i,j}(\theta));\;s\in[s_{i,j-1},s_{i,j}]\}$. Then,
$$\ds P(N_b=0|X,\theta)=\exp\left\{-\int_{s_{i,j-1}}^{s_{i,j}}\left(\phi(X_s;\theta)-\phi_{i,j,l}(\theta)\right)ds\right\}.$$
\end{proposition}

\begin{lemma}\label{POD2}
The density of $\ds (\tilde{X}_i,X^{(i)},\mathcal{L}_i,\Xi_{i})$ conditional on $(X_{\tau_{i}},X_{\tau_{i+1}},N,\theta)$, for $i=1,\ldots,m-1$, with respect to the dominating measure $\mathbb{Q}_{1,i}=\mathbb{L}^{n_i}\otimes\tilde{\mathbb{W}}_i\otimes\Xi_{i}^+$, is given by
\begin{eqnarray}\label{pod1a}
\ds &&\frac{d\mathbb{P}}{d\mathbb{Q}_{1,i}}(\tilde{X}_{i},X^{(i)},\mathcal{L}_{i},\Xi_{i}|N,X_{\tau_i},X_{\tau_{i+1}})= \nonumber\\
&=&\pi_{\tilde{\mathbb{W}_{i}^*}}(\tilde{X}_{i})
\exp\left\{\sum_{j=1}^{n_i+1}\Delta_{i,j}(1-r_{i,j}(\theta))\right\}\prod_{j=1}^{n_i+1}r_{i,j}(\theta)^{\kappa_{i,j}}\exp\left\{-\sum_{j=1}^{n_i+1}\Delta_{i,j}(\phi_{i,j,l}-m)(\theta)\right\} \nonumber \\
&\times&\prod_{j=1}^{n_i+1}\prod_{k=1}^{\kappa_{i,j}}\mathbb{I}\left[\frac{\phi(X_{\psi_{i,j,k}};\theta)-\phi_{i,j,l}(\theta)}{r_{i,j}(\theta)}<\upsilon_{i,j,k}\right]\frac{1}{a_i(X_{\tau_{i}},X_{\tau_{i+1}};\theta)},
\end{eqnarray}
where $\ds \pi_{\tilde{\mathbb{W}_{i}^*}}(\tilde{X}_{i})$ is given by (\ref{densXtj}) for intervals with events from $N$, and is 1, otherwise. Also,
\begin{equation}\label{aiRS}
a_i(X_{\tau_{i}},X_{\tau_{i+1}};\theta)=\mathbb{E}_{\tilde{\mathbb{W}_{i}^*}}\left[\exp\left\{-\int_{\tau_i}^{\tau_{i+1}} \left(\phi(X_s;\theta)-m(\theta)\right)ds\right\}\right].
\end{equation}
\end{lemma}
For $i=0$, we replace $\tilde{X}_i$ by $X_0$ and use the dominating measure $\mathbb{L}\otimes\tilde{\mathbb{W}}_0\otimes\Xi_{0}^+$. For $i=m$, we replace $\tilde{X}_i$ by $X_T$ and use the dominating measure $\mathbb{L}\otimes\tilde{\mathbb{W}}_m\otimes\Xi_{m}^+$.

\newpage

\section*{Appendix B - Simulated examples}

Simulation from a DDCP model is achieved by basically combining exact simulation of diffusions with the Poisson thinning technique that simulates an inhomogeneous Poisson process with intensity $\lambda_s$ by thinning the events from a homogeneous Poisson process with intensity $\lambda^*$ - an upper bound for $\lambda_s$, $\forall s$. Exact simulation of diffusions is performed via the EA algorithm proposed in \citet{bpr06b} and \citet{bpr07}. The EA algorithm samples from the exact law of a class of diffusion processes via retrospective rejection sampling. It proposes from (biased) Brownian motion in the case of unconditional diffusions and from Brownian bridge in the case of diffusion bridges. The algorithm is directly applied to unit diffusion coefficient processes which can always be obtained, if $\sigma$ is differentiable, by applying the Lamperti transform. The acceptance probability of EA has the form $\ds \exp\left\{-\int_{0}^t\phi (X_s)ds\right\}$, where $\phi (x) = \left(\frac{\alpha^2+\alpha'}{2}\right)(x)-l$ and
$\ds l=\inf_{u\in\mathcal{X}}(\alpha^2+\alpha')(u)/2$. The decision of whether or not to accept the proposal is taken through a Poisson process in such a way that the path of $X$ only needs to be unveiled at a random finite collection of time points.

Suppose, without loss of generality, that function $g$ is unbounded. The algorithm to simulate DDCPs is as follows:\\
\\{
\begin{tabular}[!]{|l|}
\hline\\
\parbox[!]{13cm}{
\texttt{
{\bf Exact simulation of DDCPs}
\begin{enumerate}
\setlength\itemsep{-0.5em}
\item simulate $X_0$ from $f_{0}$;
\item simulate $X$ in $(0,t]$ via EA and keep the lower and upper bounds for $X$ obtained from $\mathcal{L}$;
\item obtain an upper bound $\lambda^*$ for $\lambda_s$ in $[0,t]$ using the bounds for $X$;
\item simulate a Poisson process with rate $\ds \lambda^*$ on $[0,t]$: $\ds(t_1,\ldots,t_{n^*})$;
\item simulate $X$ at times $\ds(t_1,\ldots,t_{n^*})$, from the respective BB conditional on $\mathcal{L}$;
\item keep each of the $n^*$ points with probability $\ds R(t_i)=\frac{\lambda(t_i)}{\lambda^*}$.
\end{enumerate}
}}\\  \hline
\end{tabular}}\\
Step 2 should be performed piecewise if $t$ is big, in order to get a reasonable computational cost \citep[see][]{bpr06b}. In this case, each simulated interval provides an upper bound $\lambda^*$ based on $\mathcal{L}$.

We simulate four Cox processes. For the OU-process $dX_s=\rho X_sds+dWs$, we consider $g(X_s)=\exp(\mu+\sigma X_s)$ (exp-OU), for $\mu=0$, $\rho=0.05$, $\sigma=0.2$; and $g(X_s)=\delta\Phi(\sigma X_s)$ (cdf-OU), for $\rho=0.05$, $\sigma=0.2$, $\delta=3$.
For the (transformed) DW process $dX_s = -\rho X_s(\sigma^2X_{s}^2-\mu)ds + dW_s$, $g(X_s)=\exp(\delta + \sigma X_s)$ (exp-DW), for $\delta=0$, $\mu=1.5$, $\rho=0.05$, $\sigma=0.15$; and $g(X_s)=\delta\Phi(\sigma X_s)$ (cdf-DW), for $\mu=0.5$, $\rho=0.1$, $\sigma=0.2$, $\delta=3$.

\begin{figure}[!h]\centering
\includegraphics[scale=0.35]{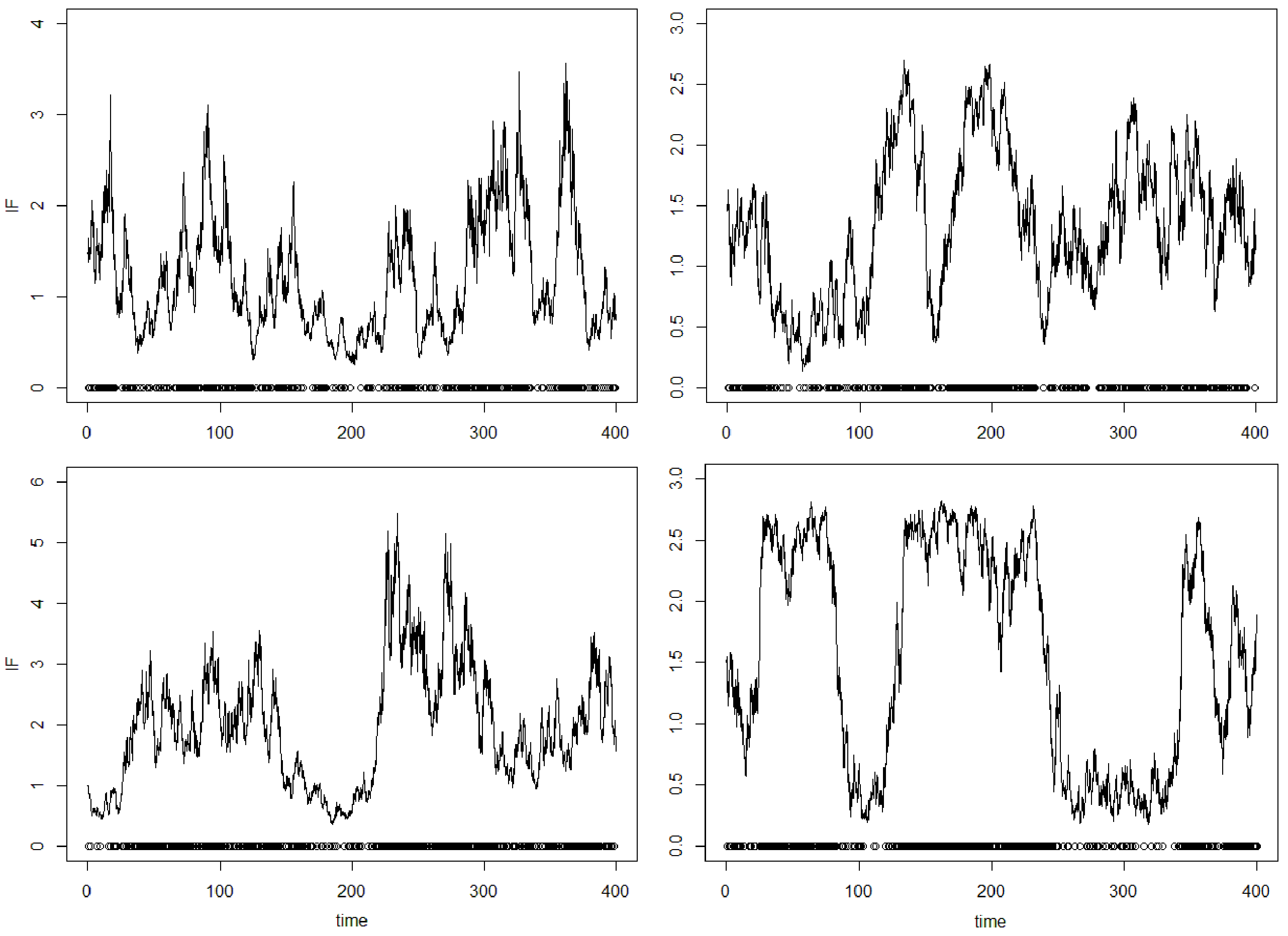}
\caption{Realization of processes exp-OU (top-left), cdf-OU (top-right), exp-DW (bottom-left) and cdf-DW (bottom-right). Number of Poisson events are 499, 518, 714 and 645, respectively.}\label{figsim1}
\end{figure}

Figure \ref{figsim1} shows one realization of each of the four processes, with $T=400$. We also compute Monte Carlo estimates of the expectation of some functionals of the processes, which are presented in Table \ref{tabsim1}.

\begin{table}[!h]\centering
\caption{Monte Carlo estimates of the expectation of some functionals of the simulated processes. An i.i.d. sample of size 50k is used. $N_T$ is the number of Poisson events, $t_{(10)}$ is the time of occurrence of the $10$-th event.}
{
\begin{tabular}{|c|c|ccccccc|}
  \hline
                        &  & 1\% & 25\% & 50\% & 75\% & 99\% & mean & s.d. \\ \hline
  \multirow{2}{*}{exp-OU}   & $N_T$ & 294 & 415 & 477 & 549 & 773 & 487.34 & 102.74 \\
                        & $t_{(10)}$ & 1.83 & 5.91 & 9.32 & 14.24 & 34.01 & 10.95 & 6.93 \\ \hline
  \multirow{2}{*}{cdf-OU} & $N_T$ & 410 & 543 & 599 & 656 & 792 & 599.58 & 82.90 \\
                        & $t_{(10)}$ & 2.14 & 4.64 & 6.61 & 9.84 & 28.06 & 8.10 & 5.31 \\  \hline
  \multirow{2}{*}{exp-DW} & $N_T$ & 118 & 167 & 654 & 1171 & 1408 & 681.20 & 469.68 \\
                        & $t_{(10)}$ & 3.70 & 7.08 & 9.40 & 13.30 & 38.12 & 11.43 & 6.89 \\ \hline
  \multirow{2}{*}{cdf-DW} & $N_T$ & 144 & 218 & 602 & 976 & 1084 & 601.20 & 349.69 \\
                        & $t_{(10)}$ & 2.66 & 5.05 & 6.50 & 8.55 & 25.40 & 7.49 & 4.19 \\  \hline
\end{tabular}}
\label{tabsim1}
\end{table}

\newpage

{\color{white} .}

\newpage

\section*{Appendix C - Sampling from a biased Brownian bridge}

We present a collection of algorithms to sample from the density
\begin{equation}\label{densXtj}
\pi_{\tilde{\mathbb{W}}_{i}^*}(\tilde{X}_i;\theta)=\frac{1}{c_i(\theta)}
\pi_{\tilde{\mathbb{W}}_i}(\tilde{X}_i)\times
\prod_{j=1}^{n_i}g(X_{s_{i,j}};\theta).
\end{equation}

Suppose that the (normal) distribution of $\tilde{X}$ under a $BB(X_{\tau_1},\tau_1,X_{\tau_2},\tau_2)$ has mean vector $\mu$ and covariance matrix $\Sigma$. Now define $\mu_0$ to be a $L$-dimensional vector with all entries equal to $b$, such that $g(u)\leq\exp(bu+c)$.
Sampling from (\ref{densXtj}) can be performed via rejection sampling by proposing a value $x$ from the distribution
\begin{equation}\label{bbbaa}
\ds \mathcal{N}\left(m,C\right),\;\;C^{-1}=\Sigma^{-1},\;\;m=C(\Sigma^{-1}\mu+\mu_0),
\end{equation}
and accepting with probability
\begin{equation*}\label{bbbb}
\ds \alpha=\prod_{l=1}^L\frac{g(x;\theta)}{\exp(bx+c)}.
\end{equation*}

If some additional conditions are satisfied though, more efficient algorithms can be used to simulate from (\ref{densXtj}). We present some examples below

\begin{itemize}

  \item {\bf Sampling from $\tilde{\mathbb{W}}^*$ when $g(u;\theta)\propto\exp\{-au^2+bu\}$, for $a\geq0$, $b\in\mathds{R}$.}

This implies that the density in (\ref{densXtj}) is a multivariate normal. Suppose first that $a>0$, then the distribution of $\tilde{X}$ under $\tilde{\mathbb{W}}^*$ is
\begin{equation}\label{bbba}
\ds \mathcal{N}\left(m,C\right),\;\;C^{-1}=\Sigma^{-1}+\Sigma_{0},\;\;m=C(\Sigma^{-1}\mu+\mu_0),
\end{equation}
where $\Sigma_0=(2a)I_L$, for $I_L$ being the $L$-dimensional identity matrix.

If $a=0$, the distribution of $\tilde{X}$ under $\tilde{\mathbb{W}}^*$ is given by (\ref{bbbaa}).

  \item {\bf Sampling from $\tilde{\mathbb{W}}^*$ when $g(u;\theta)\propto\Phi\{au+b\}$, for $(a,b)\in\mathds{R}^2$.}

This implies that the distribution of $\tilde{X}$ belongs to a general class of skew-normal distributions from which directly simulation is not feasible. Two options are available here. The first is a rejection sampling that proposes from $\mathcal{N}(\mu,\Sigma)$ and accepts with probability
\begin{equation}\label{bbbsn}
\ds \alpha=\prod_{l=1}^L \Phi(ax+b). \nonumber
\end{equation}
This algorithm has a global acceptance probability equals to $\mathbb{E}[\prod_{l=1}^L\Phi(ax+b)]$.

If the first algorithm is not efficient, simulation from (\ref{densXtj}) may be performed by considering an auxiliary embedded Gibbs sampling algorithm, as proposed by \citet{GG}. Define $W=aI_L$ and $\delta$ as the $L$-dimensional column vector with all entries equal to $-b$. Also define $\Gamma=I_L+W\Sigma W'$, $\Delta'=W\Sigma$ and $\gamma=W\mu$. Furthermore, let $A$ be the lower diagonal matrix obtained from the Cholesky decomposition
of $\Gamma$, i.e. $\Gamma=AA'$. Finally, define the region $B=\{x\in\mathds{R}^L;\;Ax>-(\gamma+\delta)\}$ and the $L$-dimensional random vectors $U_{0}^*$, $U_0$ and $U_1$. The following algorithm outputs an exact draw from (\ref{densXtj}).
\\{
\begin{tabular}[!]{|l|}
\hline
\parbox[!]{13cm}{
\texttt{
\begin{enumerate}
\setlength\itemsep{-0.5em}
\item Simulate $(U_{0}^*|U_{0}^*\in B)$, where $U_{0}^*\sim\mathcal{N}(0,I_L)$;
\item obtain $U_0=AU_{0}^*$;
\item simulate $(U_1|U_0)\sim\mathcal{N}(\Delta\Gamma^{-1}U_0,\Sigma-\Delta\Gamma^{-1}\Delta')$;
\item output $U_1+\mu$.
\end{enumerate}
}}\\  \hline
\end{tabular}}\\
The only non-trivial step from the algorithm above is Step 1, in which we need to simulate from a vector of uncorrelated standard Gaussian distribution truncated to be in a region defined by linear constraints. That is achievable by a Gibbs sampler that samples each coordinate at a time from its respective univariate truncated standard normal full conditional distribution. The algorithm is efficient since the linear constraints are defined by the lower diagonal matrix $A$, which allows us to initiate the algorithm already inside the truncated region $B$. Furthermore, the dimension $L$ will typically be small, which makes the algorithm above sufficiently fast. More details about the simulation of this general class of skew normal distributions can be found in \citet{GG}.

  \item {\bf Sampling from $\tilde{\mathbb{W}}^*$ when $g(u;\theta)$ is bounded by $\exp\{-au^2+bu+c\}$, $\forall u\in\mathds{R}$, for $a>0$, $(b,c)\in\mathds{R}^2$.}

We sample from (\ref{densXtj}) via rejection sampling by proposing a value $x$ from (\ref{bbba}) and accepting with probability
\begin{equation}\label{bbbb}
\ds \alpha=\prod_{l=1}^L\frac{g(x;\theta)}{\exp(-ax^2+bx+c)}. \nonumber
\end{equation}

  \item {\bf Sampling from $\tilde{\mathbb{W}}^*$ when $g$ is uniformly bounded above by $M\in\mathds{R}$.}

We sample from (\ref{densXtj}) via rejection sampling by proposing a value $x$ form a $\mathcal{N}(\mu,\Sigma)$ and accepting with probability
\begin{equation}\label{bbbc}
\ds \alpha=\prod_{l=1}^L\frac{g(x;\theta)}{M}. \nonumber
\end{equation}

\end{itemize}

If direct simulation from (\ref{densXtj}) is not possible and more than one of the rejection sampling algorithms above can be applied, we choose the one with the highest global acceptance probability, which may be computed analytically or empirically.

\newpage

\section*{Appendix D - The layered Brownian bridge}

Define $\ds \phi(\cdot)=\left(g+\frac{\alpha^2+\alpha'}{2}\right)(\cdot)$ and $\ds m(\theta)=\inf_{u\in \mathcal{X}}\left\{\phi(u;\theta)\right\}$. Bounds for the $\phi$ function are obtained from bounds on the Brownian bridge proposal which, in turn, are obtained through the \emph{layered Brownian bridge} construction presented in \citet{bpr07}. We ask the reader to resort to the original reference and \citet{GRL} for details about the simulation of layers and of the process given layers. In fact, we perform the layer construction by simulating layers for standard (starting and ending in 0) Brownian bridges on the respective time lengths and making the corresponding linear transformation to recover the layers for the original bridges \citep[for details, see][Appendices E and F]{GRL}. This strategy produces tighter bounds for the diffusion path which, in turn, reduces the computational cost. Also, the parameterization considered in Theorem 2 requires the simulation of standard bridges.

We define an upper bound $r_{i,j}(X)$ for function $\phi$ in $[s_{i,j-1},s_{i,j}]$ as follows:
\begin{eqnarray}
\ds \phi_{i,j,l}(\theta)&=&\inf\left\{\phi(u;\theta);\;u\in B_{i,j}\right\}, \nonumber \label{phil}\\
    \phi_{i,j,u}(\theta)&=&\sup\left\{\phi(u;\theta);\;u\in B_{i,j}\right\}, \nonumber \label{phiu} \\
    r_{i,j}(\theta)&=&\phi_{i,j,u}(\theta)-\phi_{i,j,l}(\theta);\;u\in B_{i,j}, \nonumber \label{rj}
\end{eqnarray}
where
\begin{equation}\label{layers2}
\ds \ds B_{i,j}:=[\underline{x}_{i,j}+\underline{L}_{i,j},\overline{x}_{i,j}+\overline{L}_{i,j}],\;j=1,\ldots,n_i+1, \nonumber
\end{equation}
with $\underline{x}_{i,j}=\min\{X_{s_{i,j-1}},X_{s_{i,j}}\}$, $\overline{x}_{i,j}=\max\{X_{s_{i,j-1}},X_{s_{i,j}}\}$, $\underline{L}_{ij}$ and $\overline{L}_{ij}$ being lower and upper bounds for the independent standard BBs in $(s_{i,j-1},s_{i,j})$.

For the parametrization considered in Theorem 2 (Section 5.1), we redefine:
\begin{equation}\label{layers2}
\ds \ds B_{i,j}:=[\underline{x}_{i,j}(\theta)+\underline{L}_{i,j},\overline{x}_{i,j}(\theta)+\overline{L}_{i,j}],\;j=1,\ldots,n_i+1, \nonumber
\end{equation}
with $\underline{x}_{i,j}(\theta)=\min\{X_{s_{i,j-1}}(\theta),X_{s_{i,j}}(\theta)\}$, $\overline{x}_{i,j}(\theta)=\max\{X_{s_{i,j-1}}(\theta),X_{s_{i,j}}(\theta)\}$.

\newpage

\section*{Appendix E - Sampling the partition $\tau$}

The random partition $\tau$ is updated on every iteration of the Gibbs sampler by setting a time length $\varepsilon$ and doing as follows:
\begin{enumerate}[i)]
  \item $\ds \tau_1\sim\mathcal{U}(0,\min\{\varepsilon,t_1\})$;
  \item $\ds \tau_2=\tau_1+\varepsilon$, if $t_1>\varepsilon$ and $\ds \tau_2=\tau_1+\mathcal{U}(0,\varepsilon)$, otherwise;
  \item $\ds \tau_i=\tau_1+(i-1)\varepsilon$, for $l=3,\ldots,m-1$;
  \item $\ds \tau_m=\tau_{m-1}+\varepsilon$, if $\tau_m\in(t_n,T)$ and $t_n+\varepsilon>T$ for some $m$; and $\ds \tau_m=\mathcal{U}(t_n,T)$, otherwise.
\end{enumerate}
The idea behind the algorithm to sample $\tau_2$ is that this and all the subsequent $\tau_i$ times are guaranteed to be randomly chosen in intervals of length $\varepsilon$, for a suitable choice of $\varepsilon$. Note that, if $t_1<\varepsilon$ and $\tau_2=\tau_1+\varepsilon$, the $\tau_i$ times would be restricted to intervals of length $t_1$. More specifically, if $t_1<\varepsilon$ and $\tau_2=\tau_1+\varepsilon$, we have that, marginally, $\tau_i\sim U((i-1)\varepsilon,(i-1)\varepsilon+t_1)$, for $i=2,\ldots,m-1$. This can seriously compromise the mixing of the MCMC if $t_1$ is too small.

\newpage

\section*{Appendix F - Proofs}

\subsection*{Proof of Proposition 1}

Bayes Theorem gives
\begin{eqnarray}\label{fcX2}
\ds \frac{d\tilde{\mathbb{P}}}{d\tilde{\mathbb{W}}}(X_{-\tau})&=&\frac{d\mathbb{P}}{d\mathbb{W}}(N|X,\theta)\left[\frac{d\mathbb{P}}{d\mathbb{W}}(X_{-\tau}|X_{\tau},\theta)\right]\frac{d\mathbb{P}}{d\mathbb{W}}(X_{\tau}|\theta)\frac{d\mathbb{W}}{d\mathbb{P}}(X_{\tau},N|\theta) \nonumber \\
&=&e^{T}L(X,\theta)\left[\mathcal{G}(X,\theta)\exp\left\{A(X_T;\theta)-A(X_0;\theta)\right\}\frac{d\mathbb{W}}{d\mathbb{P}}(X_{\tau}|\theta)\right]\frac{d\mathbb{P}}{d\mathbb{W}}(X_{\tau}|\theta)\frac{d\mathbb{W}}{d\mathbb{P}}(X_{\tau},N|\theta)
\nonumber \\
&=&e^{T}L(X,\theta)\mathcal{G}(X,\theta)\exp\left\{A(X_T;\theta)-A(X_0;\theta)\right\}\frac{d\mathbb{W}}{d\mathbb{P}}(X_{\tau},N|\theta). \nonumber
\end{eqnarray}
the first equality uses Girsanov's Theorem \citep[see][]{bpr06b}.

\subsection*{Proof of Proposition 2}

From Bayes Theorem
\begin{eqnarray}\label{fcext1}
\ds \frac{d\tilde{\mathbb{W}}}{d\tilde{\mathbb{W}}^*}(X_{-\tau}) &=&  \frac{d\tilde{\mathbb{W}}_0}{d\tilde{\mathbb{W}}^*}(X^{(0)}) \left[\prod_{i=1}^{m-1}\frac{d\tilde{\mathbb{W}}}{d\tilde{\mathbb{W}}^*}(X^{(i)})\right] \frac{d\tilde{\mathbb{W}}}{d\tilde{\mathbb{W}}^*}(X^{(m)}) \nonumber \\
&=& c_0(\theta)\exp\left\{A(X_0;\theta)\right\}\left[\prod_{i=1}^{m-1}\frac{c_i(\theta)}{\prod_{j=1}^{n_i}g(X_{s_{i,j}};\theta)}\right]c_m(\theta)\exp\left\{-A(X_T;\theta)\right\}\nonumber \\
&=&c(\theta) \exp\left\{A(X_0;\theta)-A(X_T;\theta)\right\} \frac{1}{\prod_{j=1}^{n}g(X_{t_j};\theta)}. \nonumber
\end{eqnarray}

\subsection*{Proof of Proposition 3}

For $i=1,\ldots,m-1$, Bayes Theorem gives that
\begin{eqnarray}\label{fcX2}
\ds \frac{d\tilde{\mathbb{P}}_i}{d\tilde{\mathbb{W}}_{i}^*}(X^{(i)}) &\propto& \frac{d\mathbb{P}_i}{d\mathbb{W}_{i}^*}(N|X,\theta)\frac{d\mathbb{P}_i}{d\mathbb{W}_{i}^*}(X^{(i)}|X_{\tau_i},X_{\tau_{i+1}},\theta) \nonumber \\
&\propto& L_i(X,\theta)\frac{d\mathbb{P}_i}{d\mathbb{W}_{i}}(X^{(i)}|X_{\tau_i},X_{\tau_{i+1}},\theta)\frac{\pi_{\tilde{\mathbb{W}}_i}}{\pi_{\tilde{\mathbb{W}}_{i}^*}}(\tilde{X}_i)\nonumber \\
&\propto& L_i(X,\theta)\mathcal{G}_i(X,\theta) \frac{1}{\prod_{j=1}^{n_i}g(X_{s_{i,j}};\theta)}\propto\exp\left\{ -\int_{\tau_{i}}^{\tau_{i+1}}\left(\phi(X_s;\theta)-m(\theta)\right)ds\right\}\leq1.\nonumber
\end{eqnarray}

For $i=0$,
\begin{eqnarray}
\ds \frac{d\tilde{\mathbb{P}}_0}{d\tilde{\mathbb{W}}_{0}^*}(X^{(0)}) &\propto& \frac{d\mathbb{P}_0}{d\mathbb{W}_{0}^*}(N|X,\theta)\frac{d\mathbb{P}_0}{d\mathbb{W}_{0}^*}(X^{(0)}|X_{\tau_{1}},\theta) \nonumber \\
&\propto& L_0(X,\theta)\frac{d\mathbb{P}_0}{d\mathbb{W}_{0}}(X^{(0)}|X_{\tau_{1}},\theta)\frac{\pi_{\tilde{\mathbb{W}}_0}}{\pi_{\tilde{\mathbb{W}}_{0}^*}}(X_0) \nonumber \\
&\propto& L_0(X,\theta)\frac{d\mathbb{P}_0}{d\mathbb{W}_{0}}(X_{\tau_{1}}|X^{(0)},\theta)\frac{d\mathbb{P}_0}{d\mathbb{W}_{0}}(X^{(0)}|\theta)e^{A(X_0;\theta)} \nonumber \\
&\propto& L_0(X,\theta)\mathcal{G}_0(X,\theta) e^{-A(X_0;\theta)}e^{A(X_0;\theta)} \nonumber \\
&\propto& L_0(X,\theta)\mathcal{G}_0(X,\theta) \propto\exp\left\{ -\int_{0}^{\tau_{1}}\left(\phi(X_s;\theta)-m(\theta)\right)ds\right\}\leq1.\nonumber
\end{eqnarray}

For $i=m$,
\begin{eqnarray}
\ds \frac{d\tilde{\mathbb{P}}_m}{d\tilde{\mathbb{W}}_{m}^*}(X^{(m)}) &\propto& \frac{d\mathbb{P}_m}{d\mathbb{W}_{m}^*}(N|X,\theta)\frac{d\mathbb{P}_m}{d\mathbb{W}_{m}^*}(X^{(m)}|X_{\tau_{m}},\theta) \nonumber \\
&\propto& L_m(X,\theta)\frac{d\mathbb{P}_m}{d\mathbb{W}_{m}}(X^{(m)}|X_{\tau_{m}},\theta)\frac{\pi_{\tilde{\mathbb{W}}_m}}{\pi_{\tilde{\mathbb{W}}_{m}^*}}(X_T) \nonumber \\
&\propto& L_m(X,\theta)\mathcal{G}_m(X,\theta) e^{A(X_T;\theta)}e^{-A(X_T;\theta)} \nonumber \\
&\propto& L_m(X,\theta)\mathcal{G}_m(X,\theta) \propto\exp\left\{ -\int_{\tau_m}^{T}\left(\phi(X_s;\theta)-m(\theta)\right)ds\right\}\leq1.\nonumber
\end{eqnarray}

\subsection*{Proof of Proposition 4}

The result comes from standard properties of Poisson processes.

\subsection*{Proof of Lemma 1}

We use the result in Proposition 3 and note that
\begin{eqnarray}\label{fcX2}
\ds \exp\left( -\int_{\tau_{i}}^{\tau_{i+1}}\left(\phi(X_s;\theta)-m(\theta)\right)ds \right) &=&
\exp\left( \sum_{j=1}^{n_i+1}\Delta_{i,j}(\phi_{i,j,l}(\theta)-m(\theta)) \right) \nonumber \\
&&\prod_{j=1}^{n_i+1}\exp\left( -\int_{s_{i,j-1}}^{s_{i,j}}\left(\phi(X_s;\theta)-\phi_{i,j,l}(\theta)\right)ds \right). \nonumber
\end{eqnarray}

\subsection*{Proof of Theorem 1}

We write the density of $(N,X,\Xi,\theta)$ w.r.t the dominating measure $\mathbb{Q}=\mathbb{Q}_1\otimes\mathbb{Q}_2\otimes\mathbb{Q}_3$, where $\mathbb{Q}_1=\otimes_{i=0}^m\mathbb{Q}_{1,i}$, where $\mathbb{Q}_{1,i}=\mathbb{L}^{n_i}\otimes\tilde{\mathbb{W}}_i\otimes\Xi_{i}^+$, for $n_0=n_m=1$, $\mathbb{Q}_2=\mathbb{N}\otimes\mathbb{L}^{m}$ and $\mathbb{Q}_3=\mathbb{L}^{d_{\theta}}$, with $d_{\theta}$ being the dimension of $\theta$. We have that
\begin{equation}\label{fctheta1}
\ds \frac{d\mathbb{P}}{d\mathbb{Q}}(N,X,\Xi,\theta) = \left[\prod_{i=0}^{m}\frac{d\mathbb{P}}{d\mathbb{Q}_{1,i}}(\tilde{X}_i,X^{(i)},\mathcal{L}_i,\Xi_{i}|X_{\tau_{i}},X_{\tau_{i+1}},N,\theta)\right]
\frac{d\mathbb{P}}{d\mathbb{Q}_2}(X_{\tau},N|\theta)\frac{d\mathbb{P}}{d\mathbb{Q}_3}(\theta).
\end{equation}
The third term on the r.h.s. of (\ref{fctheta1}) is the prior Lebesgue density of $\theta$. The first term is obtained from Lemma 1. In order to obtain the second term, we first use the chain rule for RN derivatives and the results in Propositions 1 and 2 to obtain
\begin{eqnarray}\label{fctheta2}
\ds \frac{d\tilde{\mathbb{P}}}{d\tilde{\mathbb{W}}^*}(X_{-\tau})&=&
\frac{d\tilde{\mathbb{P}}}{d\tilde{\mathbb{W}}}(X_{-\tau})
\frac{d\tilde{\mathbb{W}}}{d\tilde{\mathbb{W}}^*}(X_{-\tau}) \nonumber \\
&\overset{\theta}{\propto}&c(\theta)L(X,\theta)\mathcal{G}(X,\theta)\frac{d\mathbb{W}}{d\mathbb{P}}(X_{\tau},N|\theta)\frac{1}{\prod_{j=1}^{n}g(X_{t_j};\theta)}.
\end{eqnarray}

We take expectation on both sides of (\ref{fctheta2}) w.r.t. $\ds \tilde{\mathbb{W}}^*$ to get
\begin{equation}\label{fctheta3}
\ds \frac{d\mathbb{P}}{d\mathbb{W}}(X_{\tau},N|\theta)=c(\theta)\mathbb{E}_{\tilde{\mathbb{W}^*}}\left[L(X,\theta)\mathcal{G}(X,\theta)\frac{1}{\prod_{j=1}^{n}g(X_{t_j};\theta)}\right].
\end{equation}
Furthermore,
\begin{equation}\label{fctheta4}
\ds \frac{d\mathbb{P}}{d\mathbb{Q}_2}(X_{\tau},N|\theta)=\frac{d\mathbb{P}}{d\mathbb{W}}(X_{\tau},N|\theta)\frac{d\mathbb{W}}{d\mathbb{Q}_2}(X_{\tau},N|\theta)
\overset{\theta}{\propto}\frac{d\mathbb{P}}{d\mathbb{W}}(X_{\tau},N|\theta)\pi_{\mathbb{W}}(X_{\tau_1}|\theta),
\end{equation}
where $\ds \pi_{\mathbb{W}}(X_{\tau_1}|\theta)$ is the marginal Lebesgue density of $X_{\tau_1}$ under $\mathbb{W}$.

We now substitute (\ref{fctheta3}) into (\ref{fctheta4}) and then (\ref{fctheta4}) and expression (28) from Lemma 1 (main text) into (\ref{fctheta1}) to get
\begin{eqnarray}\label{fctheta5}
\ds \frac{d\mathbb{P}}{d\mathbb{Q}}(N,X,\Xi,\theta)&\overset{\theta}{\propto}& \pi_{\mathbb{W}}(X_0|X_{\tau_1},\theta)\pi_{\mathbb{W}}(X_{\tau_1}|\theta) \exp\left\{A(X_T;\theta)-A(X_0;\theta)-\sum_{i=0}^{m}\sum_{j=1}^{n_i+1}\Delta_{i,j}\phi_{i,j,u}(\theta)\right\} \nonumber \\
&\times&\prod_{j=1}^{n}g(X_{t_j};\theta) \prod_{i=0}^{m}\prod_{j=1}^{n_i+1}r_{i,j}(\theta)^{\kappa_{i,j}}\prod_{k=1}^{\kappa_{i,j}}\mathbb{I}\left[\frac{\phi(X_{\psi_{i,j,k}};\theta)-\phi_{i,j,l}(\theta)}{r_{i,j}(\theta)}<\upsilon_{i,j,k}\right], \nonumber
\end{eqnarray}
where
\begin{equation}\label{fctheta6}
\ds \pi_{\mathbb{W}}(X_0|X_{\tau_1},\theta)\pi_{\mathbb{W}}(X_{\tau_1}|\theta)=\frac{\pi_{\mathbb{W}}(X_{\tau_1}|X_0)f_0(X_0;\theta)}{\pi_{\mathbb{W}}(X_{\tau_1}|\theta)}\pi_{\mathbb{W}}(X_{\tau_1}|\theta)
\overset{\theta}{\propto}f_0(X_0;\theta).
\end{equation}

\subsection*{Proof of Theorem 2}

This is analogous to the proof of Theorem 1, but replacing $(X_{\tau},\tilde{X})$ by $(Y_{\tau},\tilde{Y})$ and $X\setminus(X_{\tau},\tilde{X})$ by $\omega\setminus(X_{\tau},\tilde{X})$, where $\omega$ is the linear transformation $\varphi^{-1}$ of the bridges among the $(X_{\tau},\tilde{X})$ values to make them start and end in 0. We get (for suitable dominating measures) that
\begin{eqnarray}\label{fctheta7}
\ds && \pi(N,Y_{\tau},\tilde{Y},\mathcal{L},\Xi,Y_0,\theta) \propto \nonumber \\
&&\prod_{i=0}^{m}\pi(\tilde{Y}_i,\omega_i,\mathcal{L}_i,\Xi_{i}|Y_{\tau_{i}},Y_{\tau_{i+1}},N,\theta)
\pi(X_{\tau}(\theta),N|\theta)\prod_{i=0}^{m}\eta'(Y_{\tau_{i}};\theta)\pi(\theta). \nonumber
\end{eqnarray}

The first term on the right-hand side of (\ref{fctheta7}) is obtained by adapting Lemma 1. Basically, by replacing $\pi_{\tilde{\mathbb{W}}_{i}}(\tilde{X}_i)$ with $\pi_{\tilde{\mathbb{W}}_{i}}(\tilde{X}_i(\theta);X_{\tau_{i}}(\theta),X_{\tau_{i+1}}(\theta))\prod_{j=1}^{n_i}\eta'(Y_{s_{i,j}};\theta)$,
and redefining $\phi_l$, $\phi_u$ and $m(\theta)$ as it is shown in the statement of the theorem. Furthermore, the Brownian bridge measures in the dominating measure are replaced by the measure of standard Brownian bridges (starting and ending in 0).

\newpage

\section*{Appendix G - Comparison to a discrete approximation method}\label{secsim2}

We compare the exact methodology proposed in this paper to an approximate one based on time discretization. The latter considers the Euler approximation with time step $\Delta$ for the diffusion $X$ and,
for each interval $(i\Delta,(i+1)\Delta)$, models the number of events observed in that interval as a Poisson distribution with mean $\left(\Delta\times g(X_{i\Delta};\theta)\right)$.

The MCMC algorithm for the discrete model uses the random partition approach proposed in Section 2 of the paper to update the diffusion $X$ in each sub-interval via Metropolis Hastings with a Brownian bridge proposal. The parameters are updated via MH with a Gaussian random walk proposal.

We compare the two methodologies for the exp-OU and the exp-Cauchy examples by simulating datasets from the exact (continuous time) model. We consider $T=50$ with $\mu=0$ and $\rho=0.05$ for the former, and $T=200$ with $\gamma=-1.61$ and $\sigma=0.4$ for the latter. Interval sizes of 1 unit are used in both algorithms to update the diffusion bridges for the exp-OU example. For the exp-OU, size 1 is used for the approximate method and sizes 1 or 0.25 are used for the exact method (with 0.25 being used in the pre-determined intervals where the posterior of the IF assumes higher values). The approximate method is run for different levels of discretization. For the exp-OU example we consider values of $\Delta=$ 0.1, 0.02 and 0.01. For the exp-Cauchy model, $\Delta=$ 0.05, 0.01 and 0.00625 and uniform priors on $(-3,-1.5)$ and $(0.1,1)$ are adopted for $\gamma$ and $\sigma$, respectively.

\begin{figure}[!h]\centering
\centering
\includegraphics[width=1\textwidth]{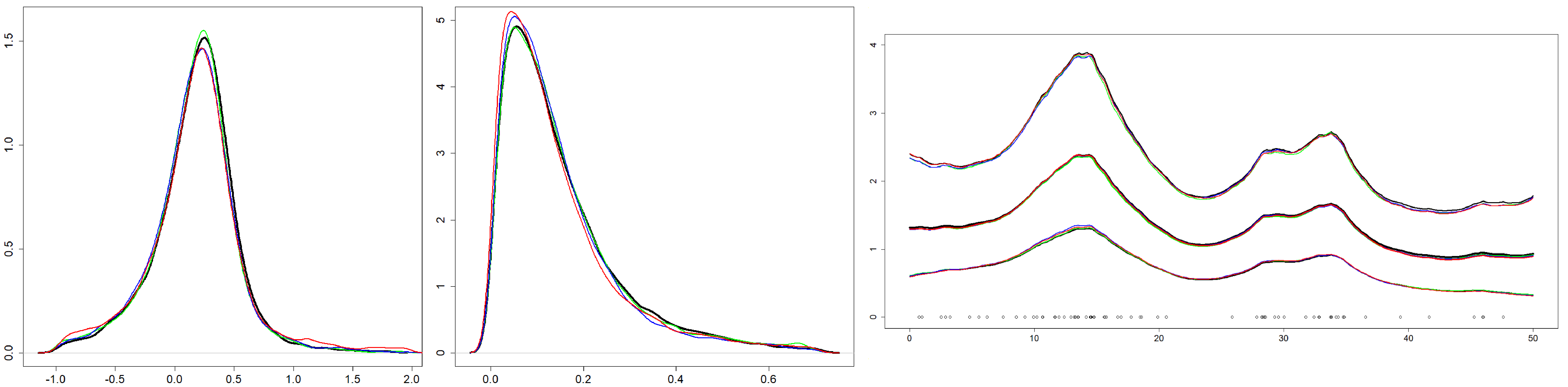}
\caption{Top: empirical posterior marginal densities of $\mu$ and $\rho$. Bottom: posterior mean and pointwise 95\% credibility interval of the IF (right). Exp-OU model with $\Delta=$ 0.1 (green), 0.02 (blue), 0.01 (red), exact (black).} \label{figcomp0}
\end{figure}

\begin{figure}[!h]\centering
\centering
\includegraphics[width=0.9\textwidth]{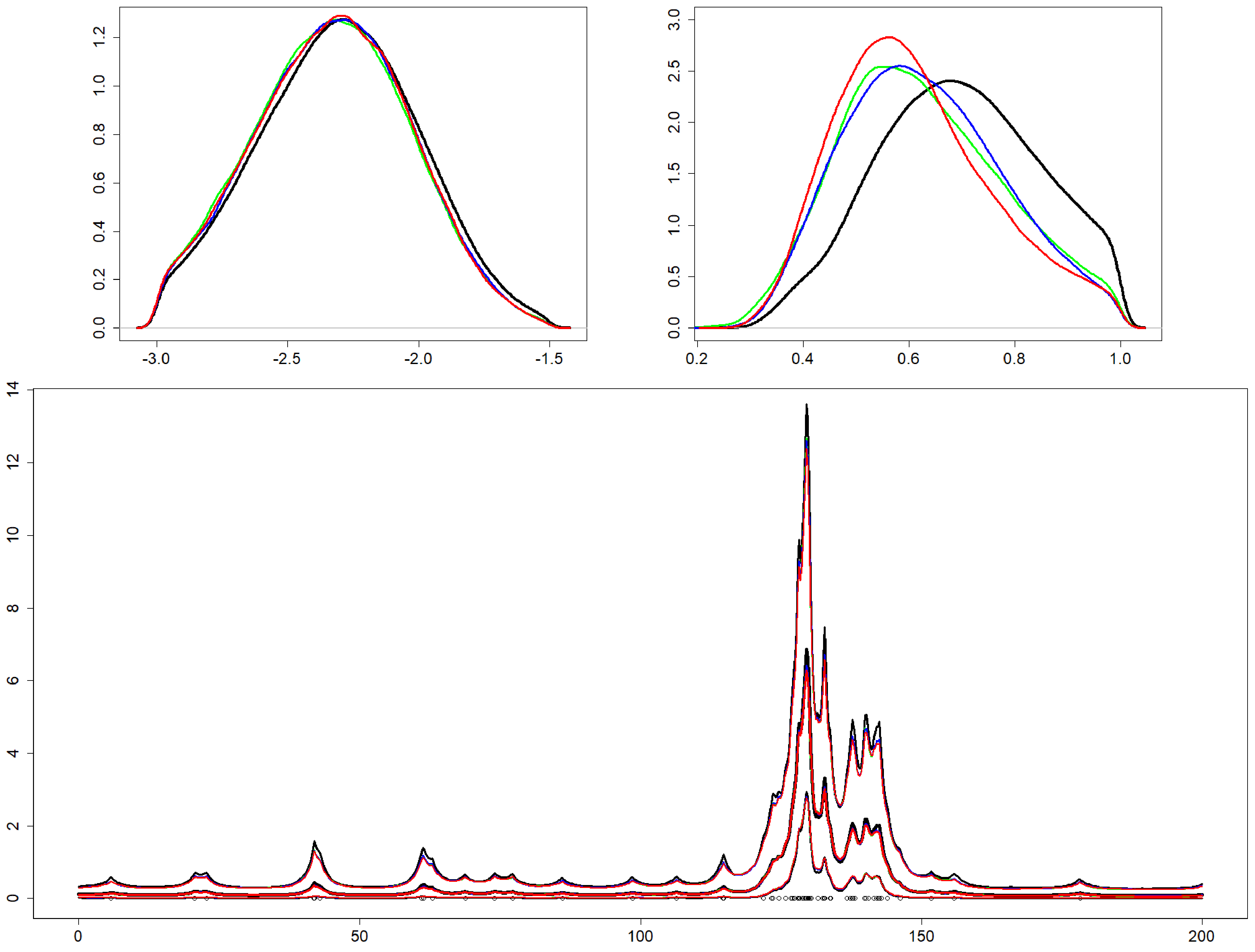}
\caption{Top: empirical posterior marginal densities of $\gamma$ and $\sigma$. Bottom: posterior mean and pointwise 95\% credibility interval of the IF. Exp-Cauchy with $\Delta=$ 0.05 (green), 0.01 (blue), 0.0065 (red), exact (black).}\label{figcomp1}
\end{figure}

Results are presented in Table \ref{tabexamp3} and Figures \ref{figcomp0} and \ref{figcomp1}. As expected, the discrete method has a lower cost to approximate the posterior for the exp-OU model when compared to the exp-Cauchy one, since the OU process is a Gaussian process. Results show a small but non-negligible (based on the respective effective sample sizes) difference between the marginal posterior densities and posterior correlation (for the finer discretization) of the parameters of the exp-OU model for the discrete and exact methods. It can be noticed that the posterior IF is well approximated by the discrete methods, since this is strongly identified (relatively to the prior) by the data.

For the exp-Cauchy example, the approximation of the posterior density of $\sigma$ is very poor even for the finer discretization, which is already twice less efficient than the exact method to sample this parameter. It is a variance parameter for which the amount of information in the data is related to (the number and distance of) the excursions of the IF away from zero. In the dataset used here, there is only one of those excursions and, considering the scale of $X$ and the posterior density of $\sigma$ obtained with the exact method, this parameter seems to be weakly identified by the data. Note, from Table 1 in the paper, how the estimates are considerably more precise when the dataset considers $T=500$ instead of $T=200$ (the prior of $\sigma$ is not truncated when $T=500$). As a consequence of the weak information about $\sigma$, the difference between the discrete and continuous models has a greater impact on the posterior, as it can be noticed in the posterior densities shown in Figure \ref{figcomp1}.

\begin{figure}[!h]\centering
\centering
\includegraphics[width=1\textwidth]{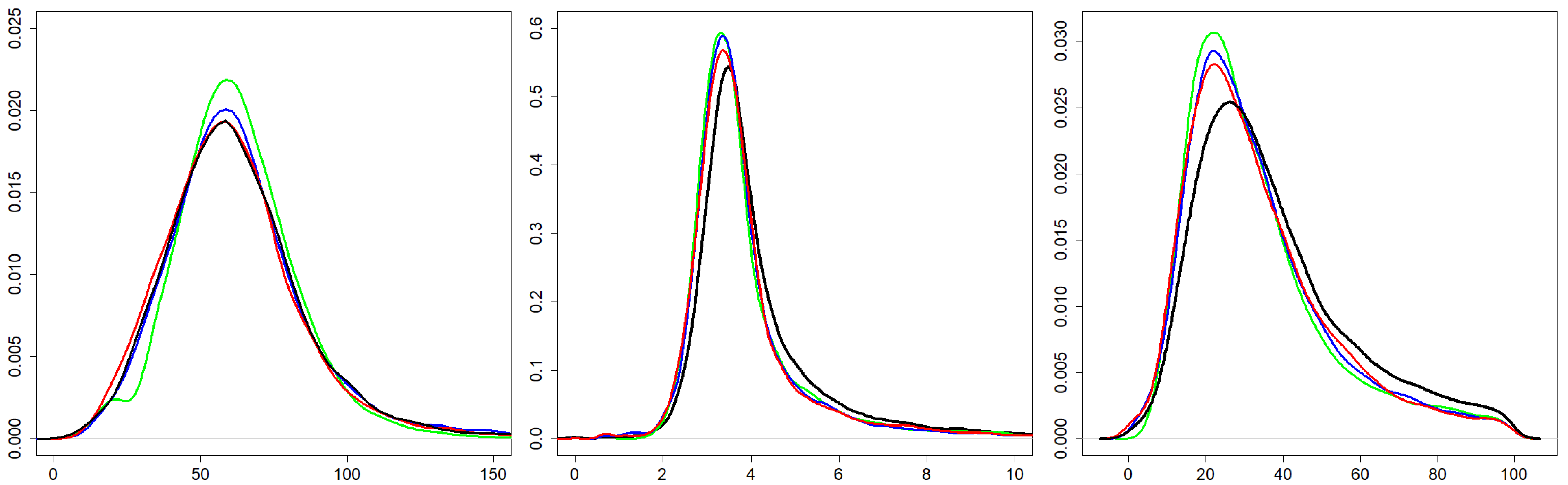}
\caption{Comparison of the predictive distribution of the (log-)number of events between the exact (in bold) and approximate methods. Exp-OU model (left): $\Delta=$ 0.1 (green), 0.02 (blue), 0.01 (red), exact (black). Exp-Cauchy for log-number (middle) and absolute number truncated at 100 - percentile 0.764 (right): $\Delta=$ 0.05 (green), 0.01 (blue), 0.0065 (red), exact (black).}\label{figcomp2}
\end{figure}

Results suggest that the discrete approximation is an impracticable option when the true diffusion model is highly non-Gaussian. Although the estimates of the IF are very similar between the two methods, the considerable difference in the parameters posterior ought to lead to considerable differences in the predictive distribution.
Figure \ref{figcomp2} compares the predictive distribution of the (log-)number of events in the future interval with the same length as the observed one. The log scale is used for the Cauchy model because of the very heavy right tail of the distributions. The truncated distribution for the absolute number is also presented, with the truncation at value 100, which is percentile 0.764. The initial value of the diffusion in the prediction interval is fixed at the posterior mean of the diffusion at the end time of the observed interval (this mean is virtually the same for the approximate and exact methods in both examples). Results show a nearly negligible bias for the exp-OU model for the two finer discretizations. For the exp-Cauchy model, the bias is still quite high for the finer discretization.

The exp-Cauchy example clearly shows the advantages of the exact methodologies when the diffusion process $X$ is highly non-Gaussian. For the exp-OU process, the analysis suggests that considerably similar results can be obtained, for compatible computational costs, for the discrete and exact methods. However, this conclusion cannot be robustly extended for different parameter configurations and data sizes. In this sense, and based on the fact that empirical evidence of convergence for the discrete method relies on obtaining results for different levels of discretization, we believe that the use of the exact method is worth even when the diffusion $X$ is a Gaussian process. Finally, the generality of the methodology proposed in this papers allows for a great variety of choices for the diffusion $X$ which, based on the results present in this session, reinforce the importance of the exact methodology for DDCPs.

\begin{table}[!h]\centering
\caption{Comparison between exact and approximate methodologies. Third row shows the posterior correlation between the estimated parameters. The last two rows show the time (in seconds) per effective sample for each parameter ($\mu$ and $\rho$ for the Exp-OU and $\gamma$ and $\sigma$ for the Exp-Cauchy).}\label{tabexamp3}
{\scriptsize
\begin{tabular}{|c|cccc|cccc|}
  \hline
                     & \multicolumn{4}{|c|}{exp-OU} & \multicolumn{4}{|c|}{exp-Cauchy} \\ \hline
            $\Delta$ & 0.1 & 0.02 & 0.01 & exact & 0.05 & 0.01 & 0.00625 & exact \\ \hline
            Corr. & 0.157 & 0.167 & 0.093 & \textbf{0.175} & -0.233 & -0.295 & -0.260 & \textbf{-0.242} \\
           time & 0.5 & 3.4 & 20.0 & \textbf{15.8} & 1.0 & 11.6 & 34.3 & \textbf{26.3} \\
           time & 0.8 & 4.9 & 23.6 & \textbf{31.1} & 8.1 & 75.5 & 225.9 & \textbf{107.6} \\  \hline
\end{tabular}}
\end{table}

\newpage

{\color{white} .}

\newpage

\bibliographystyle{chicago}
\bibliography{biblio}

\begin{thebibliography}{}

\bibitem[\protect\citeauthoryear{Beskos, Papaspiliopoulos, and Roberts}{Beskos
  et~al.}{2006}]{bpr06b}
Beskos, A., O.~Papaspiliopoulos, and G.~O. Roberts (2006).
\newblock Retrospective exact simulation of diffusion sample paths with
  applications.
\newblock {\em Bernoulli\/}~{\em 12\/}(6), 1077--1098.

\bibitem[\protect\citeauthoryear{Beskos, Papaspiliopoulos, and Roberts}{Beskos
  et~al.}{2008}]{bpr07}
Beskos, A., O.~Papaspiliopoulos, and G.~O. Roberts (2008).
\newblock A new factorisation of diffusion measure and sample path
  reconstruction.
\newblock {\em Methodology and Computing in Applied Probability\/}~{\em
  10\/}(1), 85--104.

\bibitem[\protect\citeauthoryear{Beskos, Papaspiliopoulos, Roberts, and
  Fearnhead}{Beskos et~al.}{2006}]{bpr06a}
Beskos, A., O.~Papaspiliopoulos, G.~O. Roberts, and P.~Fearnhead (2006).
\newblock Exact and computationally efficient likelihood-based inference for
  discretely observed diffusion processes (with discussion).
\newblock {\em Journal of the Royal Statistical Society, Series B\/}~{\em
  68\/}(3), 333--382.

\bibitem[\protect\citeauthoryear{Cariboni and Schoutens}{Cariboni and
  Schoutens}{2009}]{cariboni}
Cariboni, J. and W.~Schoutens (2009).
\newblock Jumps in intensity models: investigating the performance of
  {Ornstein-Uhlenbeck} processes in credit risk modeling.
\newblock {\em Metrika\/}~{\em 69}, 173--198.

\bibitem[\protect\citeauthoryear{Chib, Pitt, and Shephard}{Chib
  et~al.}{2006}]{cps}
Chib, S., M.~K. Pitt, and N.~Shephard (2006).
\newblock Likelihood based inference for diffusion driven state space models.
\newblock {\em Working paper\/}.

\bibitem[\protect\citeauthoryear{Cox}{Cox}{1955}]{Cox2}
Cox, D.~R. (1955).
\newblock Some statistical methods connected with series of events.
\newblock {\em Journal of the Royal Statistical Society, Series B\/}~{\em 17},
  129--164.

\bibitem[\protect\citeauthoryear{Dassios and Jang}{Dassios and
  Jang}{2003}]{dassios}
Dassios, A. and J.~Jang (2003).
\newblock Pricing of catastrophe reinsurance and derivatives using the cox
  process with shot noise intensity.
\newblock {\em Finance and stochastics\/}~{\em 7}, 73--95.

\bibitem[\protect\citeauthoryear{Diggle}{Diggle}{2014}]{diggle}
Diggle, P.~J. (2014).
\newblock {\em Statistical Analysis of Spatial and Spatio-Temporal Point
  Patterns\/} (3rd ed.).
\newblock London: Chapman \& Hall.

\bibitem[\protect\citeauthoryear{Doornik}{Doornik}{2007}]{Ox}
Doornik, J.~A. (2007).
\newblock {\em Object-Oriented Matrix Programming Using Ox\/} (3rd ed.).
\newblock London: Timberlake Consultants Press and Oxford.

\bibitem[\protect\citeauthoryear{Gonçalves and Gamerman}{Gonçalves and
  Gamerman}{2018}]{GG}
Gonçalves, F.~B. and D.~Gamerman (2018).
\newblock Exact {B}ayesian inference in spatiotemporal {C}ox processes driven
  by multivariate {G}aussian processes.
\newblock {\em Journal of the Royal Statistical Society - Series B\/}~{\em
  80\/}(157-175).

\bibitem[\protect\citeauthoryear{Gonçalves, {\L}atuszynski, and
  Roberts}{Gonçalves et~al.}{2023}]{GRL}
Gonçalves, F.~B., K.~{\L}atuszynski, and G.~O. Roberts (2023).
\newblock Exact {M}onte {C}arlo likelihood-based inference for jump-diffusion
  processes.
\newblock {\em To appear in Journal of the Royal Statistical Society - Series
  B\/}.

\bibitem[\protect\citeauthoryear{Iversen, Glenstrup, and Rasmussen}{Iversen
  et~al.}{2000}]{iversen}
Iversen, V.~B., A.~J. Glenstrup, and J.~Rasmussen (2000).
\newblock Internet dial-up traffic modelling.
\newblock {\em Fifteenth Nordic Teletraffic Seminar\/}.

\bibitem[\protect\citeauthoryear{Jarrett}{Jarrett}{1979}]{Jar79}
Jarrett, R.~G. (1979).
\newblock A note on the intervals between coal-mining disasters.
\newblock {\em Biometrika\/}~{\em 66}, 191--193.

\bibitem[\protect\citeauthoryear{Kloeden and Platen}{Kloeden and
  Platen}{1995}]{kloeden}
Kloeden, P. and E.~Platen (1995).
\newblock {\em Numerical Solution of Stochastic Differential Equations}.
\newblock New York: Springer.

\bibitem[\protect\citeauthoryear{{\L}atuszy{\'n}ski, Kosmidis,
  Papaspiliopoulos, and Roberts}{{\L}atuszy{\'n}ski et~al.}{2011}]{p2p}
{\L}atuszy{\'n}ski, K., I.~Kosmidis, O.~Papaspiliopoulos, and G.~Roberts
  (2011).
\newblock {Simulating events of unknown probabilities via reverse time
  martingales}.
\newblock {\em Random Structures \& Algorithms\/}~{\em 38(4)}, 441--452.

\bibitem[\protect\citeauthoryear{Lechnerová, Helisová, and
  Benes\v{s}}{Lechnerová et~al.}{2008}]{lech}
Lechnerová, R., K.~Helisová, and V.~Benes\v{s} (2008).
\newblock Cox point processes driven by {O}rnstein-{U}hlenbeck type processes.
\newblock {\em Methodology and Computing in Applied Probability\/}~{\em 10},
  315--335.

\bibitem[\protect\citeauthoryear{Legg and Chitre}{Legg and Chitre}{2012}]{legg}
Legg, M.~W. and M.~A. Chitre (2012).
\newblock Clustering of snapping shrimp snaps on long time scales: a simulation
  study.
\newblock {\em Proceedings of Acoustics\/}.

\bibitem[\protect\citeauthoryear{M{\o}ller, Syversveen, and
  Waagepetersen}{M{\o}ller et~al.}{1998}]{moller}
M{\o}ller, J., A.~R. Syversveen, and R.~P. Waagepetersen (1998).
\newblock Log {Gaussian Cox} processes.
\newblock {\em Scandinavian Journal of Statistics\/}~{\em 25}, 451--482.

\bibitem[\protect\citeauthoryear{{\O}ksendal}{{\O}ksendal}{1998}]{oksendal}
{\O}ksendal, B.~K. (1998).
\newblock {\em Stochastic Differential Equations: An Introduction with
  Applications}.
\newblock Berlin: Springer-Verlag.

\bibitem[\protect\citeauthoryear{Papaspiliopoulos, Roberts, and
  Sk\"{o}ld}{Papaspiliopoulos et~al.}{2007}]{PRSpar}
Papaspiliopoulos, O., G.~O. Roberts, and M.~Sk\"{o}ld (2007).
\newblock A general framework for the parametrization of hierarchical models.
\newblock {\em Statistical Science\/}~{\em 22}, 59--73.

\bibitem[\protect\citeauthoryear{Roberts and Sangalli}{Roberts and
  Sangalli}{2010}]{sangalli}
Roberts, G. and L.~M. Sangalli (2010).
\newblock Latent diffusion models for survival analysis.
\newblock {\em Bernoulli\/}~{\em 16}, 435--458.

\bibitem[\protect\citeauthoryear{Sermaidis, Papaspiliopoulos, Roberts, Beskos,
  and Fearnhead}{Sermaidis et~al.}{2012}]{sermai}
Sermaidis, G., O.~Papaspiliopoulos, G.~O. Roberts, A.~Beskos, and P.~Fearnhead
  (2012).
\newblock {Markov chain Monte Carlo} for exact inference for diffusions.
\newblock {\em Scandinavian Journal of Statistics\/}~{\em 40}, 294--321.

\end{thebibliography}

\end{document}